\documentclass[a4paper,12pt,oneside,final,openbib]{article}
\pdfoutput=1
\usepackage[utf8]{inputenc}
\usepackage{url}
\usepackage{cite}
\usepackage{hyperref}
\usepackage{breakurl}
\usepackage{graphicx}
\usepackage{subcaption}

\usepackage{amsmath,amssymb,epsfig}
\usepackage{color}

\begin{document}

%%%%%%%%%%%%%%%%%%%%%%%%%%%%%%%%%%%%%%%%%%%%%%%%%%%%%
\title{\bf $\mathbf{\gamma Z}$ Box at Low Energy}
\author{
Balma Duch, 
Pere Masjuan 
  \\
  {\normalsize \em
  Grup de F\'{\i}sica Te\`orica, Departament de F\'{\i}sica, 
  Universitat Aut\`onoma de Barcelona,}
  \\
  {\normalsize \em
  and Institut de F\'{\i}sica d'Altes Energies (IFAE), }
  \\
  {\normalsize \em
  The Barcelona Institute of Science and Technology (BIST), }
  \\
  {\normalsize \em
  Campus UAB, E-08193 Bellaterra (Barcelona), Spain}
\\ \\
and 
\\ \\
Hubert Spiesberger 
  \\
  {\normalsize \em 
  PRISMA$^+$ Cluster of Excellence, Institut f\"ur Kernphysik,}
  \\
  {\normalsize \em
  Johannes Gutenberg-Universit\"at, 55099 Mainz, Germany,}
}

\date{\today}

\maketitle

\begin{abstract} 
We calculate the 1-loop $\gamma Z$ box-graph correction to electron-quark 
scattering at low energy and low momentum transfer. Both electron and 
quark masses are kept non-zero. 
We express our result in terms of low-energy couplings for 
4-fermion interaction including parity-violating couplings. 
We study the zero-mass limits and show that 
a non-zero electron mass is sufficient to obtain finite, well-defined couplings 
which are insensitive to a hadronic mass cutoff. We finally discuss the 
impact of our results on the determination of the weak charge of the 
proton from polarized electron-proton scattering. 
\end{abstract}

\clearpage

%\pacs{13.35.Dx, 14.65.-q, 11.55.Hx, 12.38.Bx.}

%%%%%%%%%%%%%%%%%%%%%%%%%%%%%%%%%%%%%%%%%%%%%%%%%%%%%
\section{Introduction} 

Processes at low energies have always been playing an important 
role to study properties of the weak interaction. Atomic parity violation 
in heavy atoms, parity violation in elastic electron proton scattering, 
or Moller scattering, are outstanding examples. The Standard Model (SM) 
of the electroweak interactions provides the foundation for predictions 
of corresponding measurements, and data for the above examples 
have led to precise determinations of the weak mixing angle, one 
of the central parameters for weak interactions in the Standard Model. 
\\ 

Electroweak radiative corrections for atomic parity violation have 
been calculated for the first time in the 1980's by Marciano and Sirlin 
in Ref.~\cite{Marciano:1982mm}. Their results have been used in 
the review \cite{Erler:2013xha} where also applications to other 
low-energy measurements of the neutral-current sector of the 
Standard Model are described. Constraints on physics beyond the 
Standard Model have been obtained based on these previous 
calculations from atomic parity violation \cite{Sahoo:2021thl} and 
from the parity-violating polarization asymmetry in deep inelastic 
electron deuteron scattering at the Jefferson Lab (JLAB@6GeV) 
\cite{Wang:2014guo}. Also the interpretation of the parity-violating 
asymmetry in elastic electron-proton scattering by the Qweak 
collaboration \cite{Qweak:2018tjf} in terms of new physics refers to 
\cite{Erler:2013xha} and similar future measurements by the P2 
experiment at the MESA collider in Mainz \cite{Becker:2018ggl} 
will require as well an accurate description of electroweak radiative 
corrections. 
\\

The results of the calculation by Marciano and Sirlin in 
Ref.~\cite{Marciano:1982mm} have been collected in coupling 
coefficients, called $C_{iq}$, for 4-fermion interaction terms of electrons 
and quarks. This had been considered a convenient way to present 
one-loop radiative corrections, strictly obtained within the framework 
of the SM. Using the $C_{iq}$ in a simple tree-level calculation of the 
4-fermion interaction allows one to take into account the one-loop 
corrections of the full SM calculation. 
More recently, it has become customary to study higher-order corrections 
also in terms of a low-energy effective theory (LEFT) where the weak 
charged and neutral currents are described with the help of 4-fermion 
interaction terms in a Lagrangian. A systematic treatment would 
include the calculation of higher-order corrections both in the SM 
and the LEFT, allowing the application of renormalization group 
techniques to resum large logarithmic corrections. The complete 
matching between couplings of the SM and the LEFT at the one-loop 
level has been performed in Ref.~\cite{Dekens:2019ept} (see also 
Refs.~\cite{Hill:2019xqk,Cirigliano:2023fnz} for some applications). 
We do not follow this more general approach since we are interested 
only in the SM one-loop calculation, valid at low energies and including 
the dependence on fermion masses without any approximation. 
\\

The one-loop Feynman integrals needed for these applications 
are well-known since long. Automatic tools are available to 
generate expressions for the general case with arbitrary masses 
and kinematic variables. However, large cancellations can occur 
in some situations and a precise numerical evaluation is often 
not straightforward. It is therefore important to derive simple 
formulae with asymptotic expansions valid for large or small 
ratios of masses and kinematic variables. In particular, we found 
that the singularity structure of the box graphs renders it a 
non-trivial task to determine the correct zero-mass or zero-energy 
limits. For example, we will see below that the perturbative part 
of the singular Coulomb correction can be missed in a calculation 
where fermion masses are set to zero already at the beginning. 
\\

It is well-known that there are no fermion-mass singularities 
in the box graphs at large energy and large momentum transfer. 
This property is important for the application of the factorization 
theorem. If mass-singular terms were present in box graphs, 
they would break the factorization theorem and a universal, 
i.e.\ process-independent definition of parton distribution 
functions would not be possible. However, the results for the 
zero-energy limit given in Ref.~\cite{Marciano:1982mm}, i.e.\ in 
a kinematic regime where the factorization theorem is anyway 
not valid, do exhibit a mass logarithm. This seeming contradiction 
has motivated us to revisit the details of the box-graph calculation 
and we provide in this work a well-defined and unambigous way 
to extract the zero-energy and zero-momentum-transfer limits 
from a fully exact calculation where all masses are kept non-zero. 
We will then be able to discuss the limits for zero electron or zero 
quark mass. As a main new result we will be able to show that a 
non-zero value for an effective quark mass, or, equivalently a 
hadronic mass cutoff, is in fact not needed. Our result for the 
$\gamma Z$-box is well-defined and finite also for a zero quark 
mass, provided the electron mass is kept non-zero. This allows 
us to include a perturbative result for the $\gamma Z$-box-graph 
correction without a dependence on non-perturbative hadronic input. 
\\

The layout of the paper is as follows: in the next section~\ref{sec:notation} 
we will describe the notation and discuss some of the results for the 
$\gamma Z$-box graphs used previously. In section~\ref{sec:SMcalculation} 
we describe and discuss some details of our calculation and numerical 
results are also shown there. We end with conclusions in 
section~\ref{sec:conclusion} and relegate some additional information 
and discussion to two appendices.

%%%%%%%%%%%%%%%%%%%%%%%%%%%%%%%%%%%%%%%%%%%%%%%%%%%%%
\section{Notation and previous results}
\label{sec:notation}

The Standard Model Feynman rules for the 
\textbf{ vertices of a fermion to the photon and the $Z$-boson} 
are
%%%%%%%%%%%%%%%%%%%%%%%%
\begin{center}
\raisebox{-11mm}{
\includegraphics[width=0.12\linewidth]{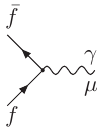}
}
$= -i e Q_f \gamma_\mu$ \quad\quad and \quad
\raisebox{-11mm}{
\includegraphics[width=0.12\linewidth]{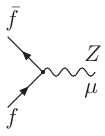}
}
$= i \displaystyle \frac{e}{2 s_W c_W} \gamma_\mu 
\left(g_V^f - g_A^f \gamma_5 \right) $
\end{center} 
%%%%%%%%%%%%%%%%%%%%%%%%
with
\begin{eqnarray}
\label{eq:boxcouplings}
g_V^f = T_f - 2 s_W^2 Q_f, 
\quad \quad 
g_A^f = T_f, 
\end{eqnarray} 
where $s_W = \sin\theta_W$ depends on the weak mixing angle, 
$c^2_W = 1 - s^2_W$, and $Q_f$ and $T_f$ are the charge 
and the 3-component of the weak isospin of fermion $f = e,\, q$. 
The normalization factor $e/(2 s_W c_W)$ of the $Z$-vertex rule 
is conveniently absorbed into the Fermi constant, 
\begin{equation}
G_F = \frac{\pi\alpha}{\sqrt{2}s_W^2 c_W^2 M_Z^2} \, , 
\label{eq:GF}
\end{equation}
with the fine structure constant $\alpha = e^2/4\pi$. 
The low-energy effective 
%Lagrangian 
\textbf{
couplings}
for weak neutral-current 
interactions of electrons and quarks, introduced in 
Ref.~\cite{Marciano:1982mm}, 
%is 
\textbf{
are defined by 
}
\begin{eqnarray} 
\label{eq:lowEL}
{\cal L}^{eq}_{\rm NC} 
&=& 
\frac{G_F}{\sqrt{2}} \sum_q \left[
C_{0q}
\bar{e} \gamma_\mu e \cdot 
\bar{q} \gamma^\mu q 
+ 
C_{1q}
\bar{e} \gamma_\mu \gamma_5 e \cdot 
\bar{q} \gamma^\mu q 
\right. 
\\[0.5ex] \nonumber
&& \quad \quad \quad \left. 
+ \,\, 
C_{2q}
\bar{e} \gamma_\mu e \cdot 
\bar{q} \gamma^\mu \gamma_5 q 
+ 
C_{3q}
\bar{e} \gamma_\mu \gamma_5 e \cdot 
\bar{q} \gamma^\mu \gamma_5 q 
\right] \, . 
\end{eqnarray} 
In Ref.~\cite{Erler:2013xha}, the normalization is expressed 
in terms of the Higgs vacuum expectation value $v$ using 
$\sqrt{2} G_F = 1/v^2$. Another convention advocated by the 
PDG~\cite{ParticleDataGroup:2024cfk} is  
\begin{equation}
C_{0q} = g_{VV}^{eq} \, , \quad
C_{1q} = g_{AV}^{eq} \, , \quad
C_{2q} = g_{VA}^{eq} \, , \quad
C_{3q} = g_{AA}^{eq} 
\end{equation}
($q = u,\, d$). The indices $V$ and $A$ refer to the vector or axial-vector 
Dirac structure of the electron and quark currents in the 4-fermion interaction 
terms of Eq.~(\ref{eq:lowEL}). At tree level in the Standard Model, 
these couplings are given by 
\begin{equation}
g_{VV}^{eq} = 2 g_V^e g_V^q \, , \quad
g_{AV}^{eq} = 2 g_A^e g_V^q \, , \quad
g_{VA}^{eq} = 2 g_V^e g_A^q \, , \quad
g_{AA}^{eq} = 2 g_A^e g_A^q \, , 
\end{equation} 
explicitly for the parity violating couplings: 
\begin{equation}
g_{AV}^{eu} = - \frac{1}{2} + \frac{4}{3} s_W^2 \, , \quad
g_{AV}^{ed} = \frac{1}{2} - \frac{2}{3} s_W^2 \, , \quad
g_{VA}^{eu} = - g_{VA}^{ed} = - \frac{1}{2} + 2 s_W^2 \, . 
\end{equation} 

Corrections to the parity-violating components $C_{1q}$, $C_{2q}$ of 
${\cal L}^{eq}_{\rm NC}$ coming from $\gamma Z$ box graphs were 
given by Marciano and Sirlin \cite{Marciano:1982mm} in the following 
form: 
\begin{eqnarray} 
\delta_{box} C_{1u} = \frac{\alpha}{2\pi} 
\left( 1 - 4 s_W^2 \right) \left(L_{M} + \frac{3}{2}\right)
\, , \quad 
\delta_{box} C_{1d} = \frac{\alpha}{2\pi} 
\frac{1}{2} \left( 1 - 4 s_W^2 \right) \left(L_{M} + \frac{3}{2}\right)\, , 
\\
\delta_{box} C_{2u} = \frac{\alpha}{2\pi} 
\left( 1 - \frac{8}{3} s_W^2 \right) \left(L_{M} + \frac{3}{2}\right)
\, , \quad 
\delta_{box} C_{2d} = \frac{\alpha}{2\pi} 
\frac{1}{2} \left( 1 - \frac{4}{3} s_W^2 \right) \left(L_{M} + \frac{3}{2}\right) \, .
\end{eqnarray} 
Restoring the charges and the vector and axial-vector couplings, 
these corrections can be combined in the following form: 
\begin{eqnarray} 
\delta_{box} C_{1q} = \delta g_{AV}^{eq}
= \frac{\alpha}{2\pi} 
Q_e Q_q g_{VA}^{eq} \, 3 \left(L_{M} + \frac{3}{2}\right)
\, , \quad 
\delta_{box} C_{2q} = \delta g_{VA}^{eq} 
= \frac{\alpha}{2\pi} 
Q_e Q_q g_{AV}^{eq} \, 3 \left(L_{M} + \frac{3}{2}\right)
\, . 
\label{eq:deltaCiMS}
\end{eqnarray} 
Here, a large logarithm of a hadronic mass scale $M$ appears, 
\begin{equation}
L_{M} = \ln \frac{M_Z^2}{M^2} \, , 
\end{equation} 
``associated with the onset of the asymptotic behavior'' (quoted 
from \cite{Marciano:1982mm}). The authors state that the constant 
3/2 was added following a calculation of Ref.~\cite{Wheater:1981ym} 
(see also the later publication \cite{Wheater:1982yk} for some 
more details), for electron scattering off a point-like proton. The 
mass scale in the logarithm was correspondingly given in the latter 
references by the proton mass. 

In Ref.~\cite{Marciano:1982mm} no details of the box-graph 
calculation are given, but that work refers to \cite{Marciano:1978ed} 
where the idea of the calculation can be found in an 
appendix\footnote{Another paper often cited in this context 
  is by Derman and Marciano \cite{Derman:1979zc}, but does 
  not contain more details of the box-graph calculation.
  }. 
%\footnote{A thorough layout of the foundation for calculations of 
%  electroweak radiative corrections for hadronic processes, 
%  like nuclear $\beta$-decays, can be found in Sirlin's 
%  classical paper \cite{Sirlin:1977sv}.
%  }
Marciano and Sanda \cite{Marciano:1978ed} 
argue that the box-graph loop integral should be calculated 
(1) taking all external momenta zero, (2) regularizing the 
singularity at zero loop momentum by a finite quark mass, 
(3) interpreting the quark mass as ``the scale at which 
quarks behave as though they were essentially free''. However, 
we argue that this prescription is not justified by a proper calculation: 
(1) it replaces the infrared divergence, due to the photon in the box 
graph becoming soft, by a mass singularity;  (2) it introduces an 
ambiguity (one could also use the electron mass as a regulator); 
(3) it creates a quark-mass singularity which is non-universal and 
therefore breaks the factorization theorem of QCD.

Erler and Su in Ref.~\cite{Erler:2013xha} quote the same result 
as shown in Eq.~(\ref{eq:deltaCiMS}) for the $\gamma Z$ box-graph 
correction to $C_{1q} = g_{AV}^{eq}$. For the other parity-violating 
coupling, $C_{2q} = g_{VA}^{eq}$, however, they use a different 
expression, namely 
\begin{eqnarray} 
\delta_{box} C_{2q} = \delta g_{VA}^{eq} 
= \frac{\alpha}{2\pi} 
Q_e Q_q g_{AV}^{eq} \, 
3 \left(\ln\frac{M_Z^2}{m_p^2} + \frac{5}{6} \right) 
\, , 
\end{eqnarray}
i.e.\ with the proton mass $m_p$ in the logarithm and a different 
constant, $5/6$ instead of $3/2$. We will see below that our 
exact calculation can tell us which of the constants is the correct 
one. 
\\

Measurements of the weak neutral current couplings at low energy 
are often described with the help of the weak charge of the nucleon. 
In particular, the measurement of the polarization asymmetry of 
elastic electron-proton scattering (at center-of-mass energy squared $s$ 
and momentum transfer $t = - Q^2$) is usually written in terms of the weak 
charge of the proton. At tree level in the Standard Model it is given by 
\begin{equation}
Q_W^{p,\text{tree}} = 1 - 4 \sin^2\!\theta_W \, , 
\label{QW_ptree_SM}
\end{equation}
and this quantity enters the prediction for the polarization asymmetry 
\begin{equation}
A^{\rm PV}=\frac{\sigma_R-\sigma_L}{\sigma_R+\sigma_L} \, . 
\label{eq:APV}
\end{equation}
Including form factor corrections \cite{Musolf:1993tb}, $F(Q^2)$, as well 
as higher-order radiative corrections~\cite{Erler:2003yk}, the prediction 
for $A^{\rm PV}$ can be written in the following way: 
\begin{equation} 
A^{\rm PV} = 
- \frac{G_FQ^2}{4\sqrt2\pi\alpha} 
\left(
Q_W^{p,\text{eff}} + \Delta_\Box(s, Q^2) + F(Q^2) 
\right) \, , 
\label{eq:APV_corr}
\end{equation} 
with \cite{Erler:2003yk}
\begin{equation} 
Q_W^{p,\text{eff}} 
= 
[\rho_{\rm NC} + \Delta_e]\left[1 - 4 \sin^2\!\hat{\theta}_W(0) + \Delta'_e\right] 
+  \Box_{WW} + \Box_{ZZ} + \Box_{\gamma Z} 
\, . 
\label{eq:QW_proton_corr}
\end{equation}
$\sin^2\!\hat{\theta}_W(0)$ is the $\overline{\mbox{MS}}$-renormalized weak 
mixing angle at scale zero. The factor $\rho_{\rm NC}$ normalizes the ratio 
of the neutral- to charged-current interaction at low energies and includes 
higher-order corrections. $\Delta_e$ and $\Delta'_e$ denote vertex and 
external leg corrections of the electron. Finally, corrections due to box graphs 
are split into their values at zero momentum transfer and zero energy plus their 
$s$- and $Q^2$-dependent parts. The constants at zero energy and $Q^2=0$ are 
denoted $\Box_{WW}$, $\Box_{ZZ}$, and $\Box_{\gamma Z}$, corresponding to 
the box-diagram contributions from $WW$, $ZZ$, and $\gamma Z$ two-boson 
exchange diagrams, respectively. Despite of the fact that the box graphs 
depend not only on properties of the nuclear target, but also on the type of 
the scattering probe, they are often considered as part of the effective weak 
charge. The $s$- and $Q^2$-dependent part of the box graphs is denoted 
$\Delta_\Box(s, Q^2)$ and added separately in Eq.~(\ref{eq:APV_corr}). 
The box-graph corrections with two heavy bosons, $WW$ and $ZZ$, can 
reliably be calculated in perturbation theory, including higher-order QCD 
corrections, see Ref.~\cite{Erler:2003yk} and contribute only to the constant 
part in the regime of low energies and low momentum transfer interesting for 
us in this work. In a naive quark-model calculation, the $\gamma Z$ box-graph 
correction for electron-proton scattering, $\Box_{\gamma Z}$ in 
Eq.~(\ref{eq:QW_proton_corr}), can be related to the box-graph corrections 
to the effective electron-quark couplings defined above:  
\begin{equation}
\square_{\gamma Z} = 
- 2 \left(2\delta_{box}C_{1u} + \delta_{box}C_{1d} \right) \, .
\label{eq:gammaZboxcont}
\end{equation}
However, the $\gamma Z$ box graphs require special care since they are 
sensitive to low-scale hadronic contributions. The calculation in 
Ref.~\cite{Erler:2003yk} assumes the form 
\begin{equation} 
\Box_{\gamma Z} = \frac{5 \hat{\alpha}}{2\pi} \left(1 - 4 \hat{s}^2 \right) 
\left[ \ln \left(\frac{M_Z^2}{\Lambda^2}\right) + C_{\gamma Z}(\Lambda)\right] \, . 
\label{eq:gammaZboxES}
\end{equation} 
Here $\hat{\alpha} = \hat{\alpha}(M_Z)$ and $\hat{s}^2 = \sin^2\!\hat{\theta}_W(M_Z)$ 
are the $\overline{\mbox{MS}}$-renormalized fine structure constant and weak 
mixing angle at the $Z$ scale, $\Lambda$ a hadronic cut-off and a constant 
$C_{\gamma Z}(\Lambda)$ whose $\Lambda$ dependence has to cancel the 
one in the logarithm. In Ref.~\cite{Erler:2003yk} it is argued that a possible choice 
is $\Lambda = m_\rho$, the $\rho$ mass, and $C_{\gamma Z}(m_\rho) = 3/2$.  
Using this expression together with $\sin^2\!\hat{\theta}_W(0)=0.23873$, i.e.\ 
the present value for the weak mixing angle from the Particle Data Group 
\cite{ParticleDataGroup:2024cfk} obtained from a Standard Model fit to all 
current experimental data, we find 
\begin{equation}
Q_W^{p,\text{eff}} = 0.0694 \, , 
\label{eq:Qwpvalue}
\end{equation}
in good agreement with the value given in \cite{Erler:2003yk} where a slightly 
different value for the weak mixing angle was used. This value should be 
kept in mind as a reference for our numerical results to be discussed below. 

The expression in Eq.~(\ref{eq:gammaZboxES}) is too simple 
to capture the full hadronic structure of the proton. An attempt to 
calculate the $\gamma Z$ 2-boson exchange in a partonic picture 
where generalized parton distribution functions are assumed to 
describe the proton internal structure, was presented in 
Ref.~\cite{Chen:2009mza}, based on earlier related work for 2-photon 
exchange corrections \cite{Chen:2004tw,Afanasev:2005mp}. 
Numerical predictions from this approach are, however, difficult 
to obtain since the required input in terms of generalized parton 
distribution functions is not well-known. In a series of papers, starting 
with Refs.~\cite{Gorchtein:2008px,Sibirtsev:2010zg,Rislow:2010vi}, 
an alternative approach based on dispersion relations has been 
developed and studied in great detail. In these references, a partonic 
picture is avoided altogether. Instead, the real part of the box-graph 
correction is related by a dispersion relation to the imaginary part and, by 
invoking the optical theorem, to the total cross section of electroproduction. 
Data are available over a large kinematic range and only little additional 
modeling is required \cite{Gorchtein:2011mz}. Most recent values for 
the $\gamma Z$ box-graph corrections relevant for the Qweak and P2 
experiments can be found in Refs.~\cite{Gorchtein:2015naa,Erler:2019rmr}. Recently, also first results with an improved uncertainty 
estimate from a lattice QCD calculation have become available in 
Ref.~\cite{Zhang:2026utt}.

Instead of following one or another of the approaches mentioned 
above, we believe that a careful analysis of the partonic one-loop 
calculation of the $\gamma Z$-box graphs is still necessary. In the 
subsequent section we are going to describe such a calculation 
and we will present a thorough study of the analytic properties of 
the result, in particular the threshold limit and the dependence on 
the fermion masses.

%%%%%%%%%%%%%%%%%%%%%%%%%%%%%%%%%%%%%%%%%%%%%%%%%%%%%
\section{Standard Model calculation with non-zero masses}
\label{sec:SMcalculation}

\begin{figure}[t!]
\centering 
    \includegraphics[width=0.8\linewidth]{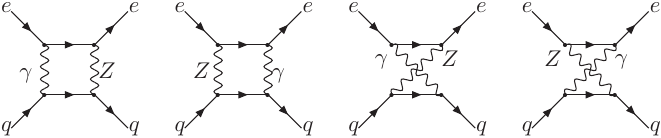}
\caption{ 
The four $\gamma Z$-box graphs. }
\label{BoxFeynDiag}
\end{figure}

We present the calculation of the one-loop $\gamma Z$-box graphs 
shown in Fig.~\ref{BoxFeynDiag} for the case of electron-quark 
scattering, $e^- q \to e^- q$, with arbitrary spins for both particles. 
The momenta of the incoming particles are denoted by $p_1$ and 
$q_1$, those of the outgoing particles by $p_2$ and $q_2$. The 
electron mass is $m$, the quark mass $M$. The Mandelstam 
variables are 
\begin{equation} 
s = (p_1+q_1)^2, \quad
t = (p_1 - p_2)^2, \quad
u = (p_1 - q_2)^2
\label{eq:mandelstam}
\end{equation} 
with $s+t+u = 2 m^2 + 2 M^2$. In order to extract the effective 
low-energy coupling constants, we have to take the threshold 
limit at zero momentum transfer, i.e.\ 
\begin{equation} 
s \to s_0 = (m + M)^2, \quad 
t \to t_0 = 0 \, . 
\label{eq:threshold}
\end{equation} 

%%%%%%%%%%%%%%%%%%%%%%%%
%%%%%%%%%%%%%%%%%%%%%%%%

To identify the $\gamma Z$-box-graph corrections for the low-energy 
couplings, $\delta_{box}C_{iq}$, we calculate the interference with 
the one-photon exchange tree graph and keep track of the dependence 
on the spins of the incoming particles, i.e.\ with spin vectors $s_e$ 
and $s_q$ for the incoming electron and quark, respectively. Using 
the effective Lagrangian defined in Eq.~(\ref{eq:lowEL}) and 
distinguishing the cases where only the electron is polarized 
(Eq.~(\ref{epolarized})), only the quark is polarized (Eq.~(\ref{qpolarized})), 
both are polarized (Eq.~(\ref{doublepolarized})), and neither is polarized 
(Eq.~(\ref{unpolarized})), we find: 
\begin{align}
\frac{2 \sqrt{2} e^{2} G_{F} m}{t} &\left[
(q_{2} \cdot s_{e}) \left(
\delta_{box}C_{1q} (2M^{2} + 2m^{2} - 2s - t) 
+ 2 \delta_{box}C_{2q} t 
\right) \right. \notag \\
&\left. + \; 
(p_{2} \cdot s_{e}) \left(
\delta_{box}C_{1q} (m^{2} + 3M^{2} - s) 
+ \delta_{box}C_{2q} t 
\right)
\right] \, ,
\label{epolarized} 
\\
\frac{2 \sqrt{2} e^{2} G_{F} M}{t} &\left[
(q_{2} \cdot s_{q}) \left(
\delta_{box}C_{2q} (3m^{2} + M^{2} - s) 
+ \delta_{box}C_{1q} t 
\right) \right. \notag \\
&\left. + \;
(p_{2} \cdot s_{q}) \left(
2 \delta_{box}C_{1q} t 
+ \delta_{box}C_{2q} (2m^{2} + 2M^{2} - 2s - t)
\right)
\right] \, , 
\label{qpolarized} 
\\
\frac{4 \sqrt{2} e^{2} G_{F} m M}{t} &\left[
(s_{e} \cdot s_{q}) \left(
\delta_{box}C_{0q} t + \delta_{box}C_{3q} (m^{2} + M^{2} - s)
\right) \right. \notag \\
& + \;
(s_{e} \cdot p_{2}) (s_{q} \cdot p_{2}) \; \delta_{box}C_{3q} 
+ (s_{e} \cdot p_{2}) (s_{q} \cdot q_{2}) (\delta_{box}C_{0q} + \delta_{box}C_{3q}) \notag \\
&\left. + \;
(s_{e} \cdot q_{2} ) (s_{q} \cdot q_{2}) \; \delta_{box}C_{3q}
\right] \, , 
\label{doublepolarized}
\\
\frac{-\sqrt{2} e^{2} G_{F}}{t} &\left[
\delta_{box}C_{0q} \left(2(m^{2} + M^{2} - s)^{2} + 2st + t^{2}\right) 
+ \delta_{box}C_{3q} t \left(2(m^{2} + M^{2} - s) - t\right)
\right] \, .
\label{unpolarized} 
\end{align}
Similarly, we calculate the interference term with the $\gamma Z$ 
box in the Standard Model. To perform the computation, we use FeynCalc 
\cite{Shtabovenko:2023idz,Shtabovenko:2020gxv,Mertig:1990an} 
and reduce the one-loop integrals to the well-known Passarino-Veltman 
scalar integrals denoted by $A_0$, $B_0$, $C_0$, and $D_0$ 
\cite{Passarino:1978jh,tHooft:1978jhc}. Their definition including conventions for 
their arguments and normalization can be found in the FeynCalc documentation 
\cite{Shtabovenko:2023idz,Shtabovenko:2020gxv,Mertig:1990an}. 
We also found Package-X \cite{Patel:2016fam} and FeynHelpers 
\cite{Shtabovenko:2016whf} useful for the calculation. We emphasize 
that the $Z$-boson mass is always kept finite. The result is therefore 
ultraviolet finite. We have checked that taking the limit $M_Z \to \infty$ 
introduces small errors, in the order of 1 to 2\,\% for energies 
of interest in our study. In this limit we could also compare our result 
with a corresponding calculation for the $s$-channel process 
$e^+ e^- \to \mu^+ \mu^-$ described in Ref.~\cite{Kollatzsch:2022bqa}. 
We should mention that the result contains infrared divergences if 
the momentum transfer is non-zero, $t \neq 0$. This divergence is 
regularized in dimensional regularization, but disappears at $t=0$. 

 Since the box-graph corrections are ultraviolet finite, they are 
also renormalization scheme independent. They are also independent 
of the choice of the gauge. These properties are  obvious from 
the fact that the box graphs come with a specific combination of 
charge and coupling constants which does not appear in other 
one-loop corrections. The other one-loop corrections from self-energy 
graphs and vertex corrections can therefore be taken from an 
independent calculation for which we refer to Ref.~\cite{Marciano:1982mm}. 
For completeness we mention that the one-loop corrections in 
Ref.~\cite{Marciano:1982mm} are expressed in terms of the 
muon decay constant and the running renormalized weak mixing 
angle in the $\overline{\rm MS}$ scheme.

The resulting expression keeping track of the spin dependence can 
be written in the following way: 
\begin{equation}
\textup{Re} \left({\cal M}_\gamma^\ast {\cal M}_{\gamma Z} \right)
= 
A + 
B \,(p_2 \cdot s_e) + 
C \,(q_2 \cdot s_e) + 
D \,(s_e \cdot s_q) + \ldots 
\end{equation} 
where ${\cal M}_\gamma$ and ${\cal M}_{\gamma Z}$ are the 
matrix elements for the leading-order one-photon exchange and 
the one-loop $\gamma Z$-box graphs, respectively.  
The first term, $A$, has no spin-dependence, $B$ and $C$ are 
responsible for the single-spin dependence of the electron and 
$D$ is one of the coefficients which depends on both spin vectors. 
These coefficients are functions of the kinematic variables, for 
example $s$ and $t$. Other terms could be used as well and we 
have checked that we find the same result using terms containing 
the quark-spin vector instead. Identifying the two calculations, within 
the Standard Model and with the effective low-energy Lagrangian, 
one obtains a set of equations from which one can determine the 
box-graph corrections: 
\begin{align}
\delta_{box} \hat{C}_{1q}(s,t) = &
\frac{t(C - 2B)}{2\sqrt{2} e^2 G_F m (4M^2 + t)} 
\, , 
\label{C1AB} 
\\
\delta_{box} \hat{C}_{2q}(s,t) = &
\frac{-B(-2m^2 - 2M^2 + 2s + t) - C(m^2 + 3M^2 - s)}{2\sqrt{2}e^2 G_F m (4M^2 + t)}
\, , 
\label{C2AB}
\\
\delta_{box} \hat{C}_{0q}(s,t) = & 
\frac{t}{4\sqrt{2}e^2 G_F m M} 
\frac{D(2(M^2+m^2-s)-t)t + 4 A m M (m^2+M^2-s)}{2(m^2+M^2-s)^3+2(m^2+M^2-s)s t-(m^2+M^2-s)t^2+t^3} 
\, , 
\label{C0CD} 
\\
\delta_{box} \hat{C}_{3q}(s,t) = & 
\frac{-t}{4\sqrt{2}e^2 G_F m M} 
\frac{D(2(M^2+m^2-s)^2+2st+t^2) + 4 m M A t}{2(m^2+M^2-s)^3+2(m^2+M^2-s)s t-(m^2+M^2-s)t^2+t^3} 
\, . 
\label{C3CD}
\end{align}
We have denoted the results of this calculation by $\delta_{box} \hat{C}_{iq}(s,t)$ 
since one can keep the full dependence on kinematic variables. The expressions 
are lengthy and can be found in the appendix. However, only the constant terms 
remaining after taking the threshold limit can be absorbed into the definition of 
the low-energy effective couplings.

%%%%%%%%%%%%%%%%%%%%%%%%
\begin{figure}[b!]
\centering
\begin{subfigure}[b]{0.49\textwidth}
       \begin{picture}(100,100)(0,0)
       \put(0,0){\includegraphics[width=\textwidth]{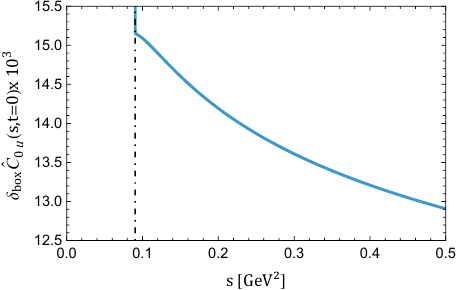}} 
       \put(125,80){\includegraphics[width=0.43\textwidth]{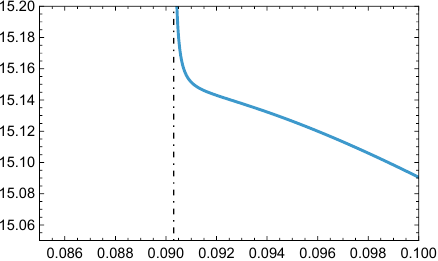}} 
       \end{picture}
        \caption{$\delta_{box}\hat{C}_{0u}(s,t=0)$}
        \vspace*{3mm}
        \label{C0s}
\end{subfigure}
%\hspace{0.5 cm}
\begin{subfigure}[b]{0.49\textwidth}
        \includegraphics[width=\textwidth]{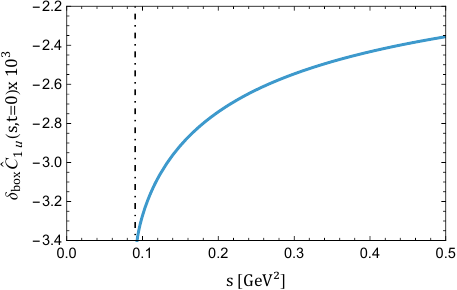}
        \caption{$\delta_{box}\hat{C}_{1u}(s,t=0)$}
        \vspace*{3mm}
        \label{C1s}
\end{subfigure}
\begin{subfigure}[b]{0.49\textwidth}
        \includegraphics[width=\textwidth]{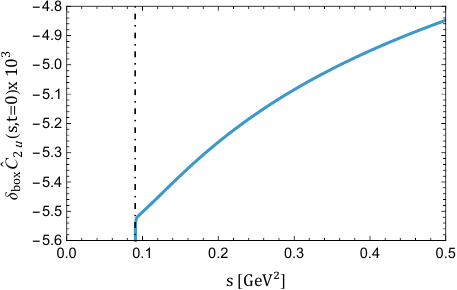}
        \caption{$\delta_{box}\hat{C}_{2u}(s,t=0)$}
        \label{C2s}
\end{subfigure}
%\hspace{0.5 cm}
\begin{subfigure}[b]{0.49\textwidth}
        \includegraphics[width=\textwidth]{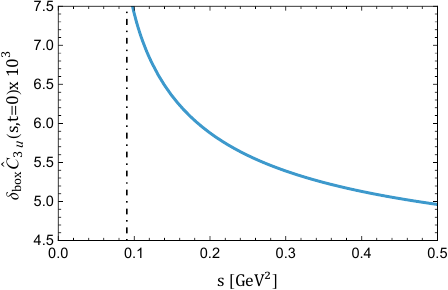}
        \caption{$\delta_{box}\hat{C}_{3u}(s,t=0)$}
        \label{C3s}
\end{subfigure}
\caption{
The box-graph corrections $\delta_{box}\hat{C}_{iu}(s, t=0)$ as a 
function of the Mandelstam variable $s$, evaluated at fixed $t = 0$ 
for a quark mass of $M = 0.3$ GeV, see Eqs.~(\ref{eq:deltaboxC0st} - 
\ref{eq:deltaboxC3st}) in the appendix. For $\delta_{box}\hat{C}_{0u}(s,t=0)$ 
we inserted a zoom-in at threshold to show that the function is in fact smooth. 
The vertical dash-dotted lines indicate the position of the threshold, 
$s_0 = (m+M)^2$. 
}
\label{C1bluered}
\end{figure}
%%%%%%%%%%%%%%%%%%%%%%%%

We find that the box-graph corrections are regular at $t=0$. However, the 
$\delta_{box} \hat{C}_{iq}(s,t=0)$ are singular at threshold, $s \to (m+M)^2$. 
Plots of $\delta_{box} \hat{C}_{iq}(s,t=0)$ as a function of $s$ for a $u$-quark 
are shown in Fig.~\ref{C1bluered}. The singularity can be traced back to the 
scalar 3-point function $C_0[m^2, M^2, s, m^2, 0, M^2]$ (or its implementaion 
in Package-X denoted {\tt ScalarC0IR6}, see the FeynCalc documentation 
\cite{Shtabovenko:2023idz,Shtabovenko:2020gxv,Mertig:1990an} and 
\cite{Patel:2016fam} for its definition). Keeping the leading singular terms only, 
we find 
\begin{align}
\delta_{box}\hat{C}_{0q}^{SE} &= 
\frac{\alpha Q_e Q_q g^{eq}_{VV} \pi \sqrt{m M}}{\sqrt{s-(M+m)^2}} \, , 
\label{C0SEterm} 
\\
\delta_{box}\hat{C}_{1q}^{SE} &= 
- \frac{\alpha Q_e Q_q g^{eq}_{AV} \pi \sqrt{m M}}{\sqrt{s-(M+m)^2}} \, , 
\label{C1SEterm} 
\\
\delta_{box}\hat{C}_{2q}^{SE} &= 
- \frac{\alpha Q_e Q_q g^{eq}_{VA} \pi \sqrt{m M}}{\sqrt{s-(M+m)^2}} \, , 
\label{C2SEterm} 
\\
\delta_{box}\hat{C}_{3q}^{SE} &= 
\frac{ \alpha Q_e Q_q g^{eq}_{AA} \pi \sqrt{m M}}{\sqrt{s-(M+m)^2}} \, . 
\label{C3SEterm}
\end{align} 
We note that these singular terms appear to be proportional to the 
square root of the masses, both $m$ and $M$. A calculation where 
one the masses is set to zero from the very beginning will 
therefore miss these terms. 

Physically, this singularity is caused by the fact that the scattering particles 
may form a bound state when the kinetic energy is low. The effect is known 
as Sommerfeld enhancement in the literature (therefore the label $SE$ in the 
above equations), see for example \cite{Sommerfeld:1931qaf,Iengo:2009ni}. 
It can be written in a simple form when we use the energy 
of one of the scattering particle, say the electron, $E_{\mathrm{lab}} \geq m$, 
in the rest frame of the other particle, i.e.\ the quark. Substituting 
$s = M^2 + m^2 + 2 M E_{\mathrm{lab}}$, the denominator in 
Eqs.~(\ref{C0SEterm} - \ref{C3SEterm}) can then be seen to be the kinetic 
energy of the electron and we can identify the factor 
\begin{equation}
S = \frac{\alpha\pi \sqrt{m M}}{\sqrt{s - (m+M)^2}} \to \frac{\alpha \pi}{\beta}
\end{equation} 
with the velocity $\beta$. This is in fact only the first-order term in a perturbative 
expansion of what was calculated originally by Sommerfeld from the modification 
of a scattering plane-wave in the Coulomb field of a nucleus with charge $Z$, 
\begin{equation}
S(v) = 
\frac{\psi \psi_n}{\psi^0 \psi_n^0} 
= 
\frac{(2\pi \alpha Z)/\beta}{1 - e^{-(2\pi \alpha Z)/\beta}}
\, . 
\end{equation} 
The terms shown in Eqs.~(\ref{C0SEterm}-\ref{C3SEterm}) are therefore 
the first-order part of what is known as Coulomb correction. To isolate the 
low-energy couplings $C_{iq}$ from these corrections, we subtract 
the velocity-dependent terms arising from box diagrams and define 
\begin{equation}
\delta_{\mathrm{box}}C_{iq}^{\mathrm{eff}}(s,t=0) 
\equiv 
\delta_{\mathrm{box}}\hat{C}_{iq}(s,t=0) - \delta_{\mathrm{box}}\hat{C}_{iq}^{SE}(s,t=0) \, . 
\end{equation}
After this subtraction one can safely take the threshold limit $s \to (m+M)^2$ 
and obtain 
\begin{equation} 
\delta_{\mathrm{box}}C_{iq} = 
\delta_{\mathrm{box}}C_{iq}^{\mathrm{eff}}(s=(m+M)^2, t=0) \, . 
\end{equation}
We believe that the above prescription is reasonable since the subtracted 
term agrees with Sommerfeld's enhancement factor. It is the only term 
which contains the factor $\alpha \pi$. 
%but from our calculation we can not 
%exclude the possibility that other finite terms proportional to $\alpha/\pi$ 
%are also associated with Coulomb corrections. A more complete study is 
%required to match the one-loop calculation with a calculation that includes 
%Coulomb corrections, for example by solving numerically the Dirac equation 
%for the electron in the field of the proton, or by applying resummation techniques. 
Comparison with Ref.~\cite{Hill:2023bfh} provides formal support for this 
statement, confirming that our procedure is justified and that, at this order, 
the Sommerfeld-enhancement fully accounts for the Coulomb corrections.

Keeping both electron and quark masses non-zero, we find at threshold: 
\begin{align}
\delta_{box}C_{0q} =& 
- \frac{\alpha}{4\pi} Q_{e}Q_{q} 
\left[ 
6 g_{VV}^{eq} \frac{m M}{m^2-M^2} \ln \left(\frac{m^2}{M^2}\right)
+ 6 g_{AA}^{eq} \left(\frac{-m^2 L_m + M^2 L_M}{m^2-M^2} - \frac{3}{2}\right)\right] 
\, , 
\label{C0thresholdgeneric}
\\
\delta_{box}C_{1q} =& 
- \frac{\alpha}{4\pi}Q_{e}Q_{q}
\Bigg[ 
g_{VA}^{eq} 
  \left(\frac{2 m^2 \left(3 m^2-5 M^2\right)}{\left(m^2-M^2\right)^2} L_m 
       - \frac{2 M^2 \left( m^2-3 M^2\right)}{\left(m^2-M^2\right)^2} L_M 
       + \frac{5 m^2-9 M^2}{m^2-M^2} 
  \right) 
\notag 
\\ & \quad \quad \quad \quad 
+ 4 g_{AV}^{eq} 
  \left(\frac{m M \left( m^2-2 M^2\right)}{\left(m^2-M^2\right)^2} 
  \ln \left(\frac{M^2}{m^2}\right) 
  - \frac{m M}{m^2 - M^2} 
  \right)
\Bigg] 
\, , 
\label{C1thresholdgeneric}
\\
\delta_{box}C_{2q} =& 
- \frac{\alpha}{4\pi}Q_{e}Q_{q}
\Bigg[
g_{AV}^{eq} 
\left(
\frac{2 m^2 \left(3 m^2-M^2\right)}{\left(m^2-M^2\right)^2} L_m 
- \frac{2 M^2 \left(5 m^2-3 M^2\right)}{\left(m^2-M^2\right)^2} L_M 
+ \frac{9 m^2-5 M^2}{m^2-M^2} 
\right) 
\notag
\\ & \quad \quad \quad \quad 
+ 4 g_{VA}^{eq} 
  \left( \frac{m M \left(2 m^2-M^2\right)}{\left(m^2-M^2\right)^2} 
     \ln \left(\frac{M^2}{m^2}\right) 
  + \frac{m M}{m^2-M^2}
  \right)
\Bigg] 
\, , 
\label{C2thresholdgeneric}
\\
\delta_{box}C_{3q} =& 
\frac{\alpha}{4\pi} Q_{e}Q_{q} 
\left[8 g_{AA}^{eq} \ln \left(\frac{m^2}{M^2}\right)\frac{m M}{M^2-m^2} 
+ 8 g_{VV}^{eq}\left(\frac{m^2 L_m - M^2 L_M}{m^2-M^2}+1\right)\right] 
\, , 
\label{C3thresholdgeneric}
\end{align} 
where we have again used the abbreviation $L_M = \ln \left(M_Z^2 / M^2\right)$ 
and now also $L_m = \ln \left(M_Z^2 / m^2\right)$.  

\begin{figure}[t!]
\centering
\begin{subfigure}[b]{0.49\textwidth}
        \includegraphics[width=\textwidth]{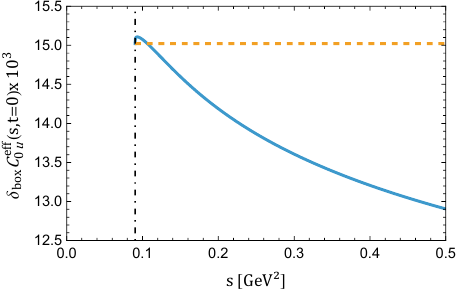}
        \caption{$\delta_{box}C_{0u}^{\mathrm{eff}}(s,t=0)$ and $\delta_{box}{C}_{0u}$}
        \vspace*{3mm}
        \label{C0sSE}
\end{subfigure}
%\hspace{0.5 cm}
\begin{subfigure}[b]{0.49\textwidth}
        \includegraphics[width=\textwidth]{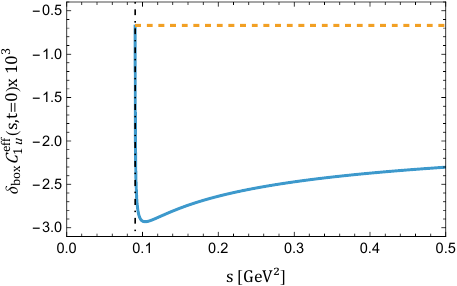}
        \caption{$\delta_{box}C_{1u}^{\mathrm{eff}}(s,t=0)$ and  $\delta_{box}{C}_{1u}$}
        \vspace*{3mm}
        \label{C1sSE}
\end{subfigure}
\begin{subfigure}[b]{0.49\textwidth}
        \includegraphics[width=\textwidth]{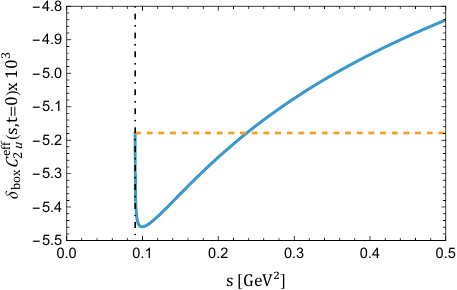}
        \caption{$\delta_{box}C_{2u}^{\mathrm{eff}}(s,t=0)$ and  $\delta_{box}{C}_{2u}$}
        \label{C2sSE}
\end{subfigure}
%\hspace{0.5 cm}
\begin{subfigure}[b]{0.47\textwidth}
        \includegraphics[width=\textwidth]{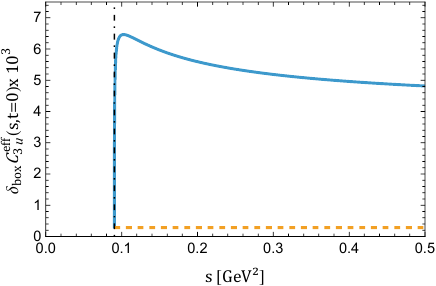}
        \caption{$\delta_{box}C_{3u}^{\mathrm{eff}}(s,t=0)$ and  $\delta_{box}{C}_{3u}$}
        \label{C3sSE}
\end{subfigure}
\caption{ 
The subtracted box-graph corrections $\delta_{box}C_{iu}^{\mathrm{eff}}$ 
as a function of the Mandelstam variable $s$, evaluated at fixed $t = 0$ for 
a quark mass of $M = 0.3$ GeV. The orange dashed lines show the value of 
$\delta_{box}C_{iu}$ at threshold ($s\xrightarrow[]{} (M+m)^2$) as 
given in Eqs.~(\ref{C0thresholdgeneric} - \ref{C3thresholdgeneric}). 
The vertical dash-dotted lines indicate the position of the threshold. 
}
\label{fig:CiuSE}
\end{figure}

\begin{figure}[h!]
\centering
\begin{subfigure}[b]{0.49\textwidth}
        \includegraphics[width=\textwidth]{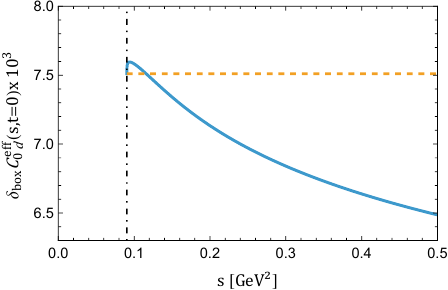}
        \caption{$\delta_{box}C_{0d}^{\mathrm{eff}}(s,t=0)$ and $\delta_{box}{C}_{0d}$}
        \vspace*{3mm}
        \label{C0sSE}
\end{subfigure}
%\hspace{0.5 cm}
\begin{subfigure}[b]{0.49\textwidth}
        \includegraphics[width=\textwidth]{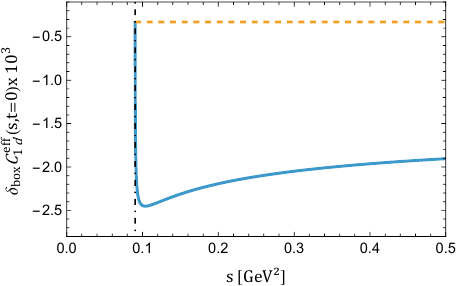}
        \caption{$\delta_{box}C_{1d}^{\mathrm{eff}}(s,t=0)$ and  $\delta_{box}{C}_{1d}$}
        \vspace*{3mm}
        \label{C1sSE}
\end{subfigure}
\begin{subfigure}[b]{0.49\textwidth}
        \includegraphics[width=\textwidth]{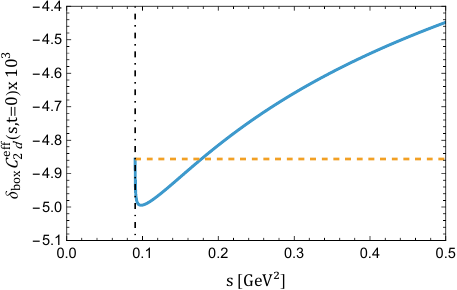}
        \caption{$\delta_{box}C_{2d}^{\mathrm{eff}}(s,t=0)$ and  $\delta_{box}{C}_{2d}$}
        \label{C2sSE}
\end{subfigure}
%\hspace{0.5 cm}
\begin{subfigure}[b]{0.49\textwidth}
        \includegraphics[width=\textwidth]{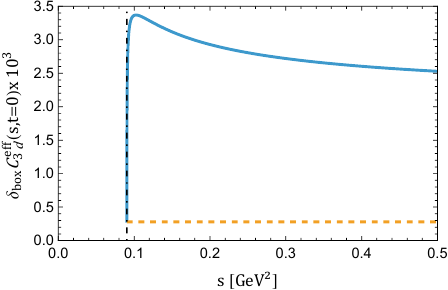}
        \caption{$\delta_{box}C_{3d}^{\mathrm{eff}}(s,t=0)$ and  $\delta_{box}{C}_{3d}$}
        \label{C3sSE}
\end{subfigure}
\caption{ 
The subtracted box-graph corrections $\delta_{box}C_{id}^{\mathrm{eff}}$ 
as in Fig.~\ref{fig:CiuSE} but now for  down quarks as a function of the 
Mandelstam variable $s$, evaluated at fixed $t = 0$ for 
a quark mass of $M = 0.3$ GeV. The orange dashed lines show the value of 
$\delta_{box}C_{iu}$ at threshold ($s\xrightarrow[]{} (M+m)^2$) as 
given in Eqs.~(\ref{C0thresholdgeneric} - \ref{C3thresholdgeneric}). 
The vertical dash-dotted lines indicate the position of the threshold. 
}
\label{fig:CidSE}
\end{figure}

Figures \ref{fig:CiuSE} and \ref{fig:CidSE} show results for the $s$-dependent 
box-graph corrections $\delta_{\mathrm{box}} C^{\mathrm{eff}}_{iq}(s,t=0)$ 
for up- and down-quarks after subtraction of the Sommerfeld terms together 
with the constant term at threshold, Eqs.~(\ref{C0thresholdgeneric} - 
\ref{C3thresholdgeneric}). 

%%%%%%%%%%%%%%%%%%%%%%%%
%%%%%%%%%%%%%%%%%%%%%%%%
%
%%%%%%%%%%%%%%%%%%%%%%%%
%%%%%%%%%%%%%%%%%%%%%%%%

We note that all expressions are regular for equal masses, $m = M$. 
Also, it is possible to take one mass to zero, but not both. The resulting 
expressions are much simpler. For zero electron mass, $m=0$, one 
finds 
\begin{align}
\delta_{box}C_{0q}(m=0) =& 
\frac{\alpha}{4\pi} Q_{e}Q_{q} \cdot 6 g_{AA}^{eq} 
\left(\ln \left(\frac{M_Z^2}{M^2}\right) + \frac{3}{2}\right)
\, , 
\label{C0thresholdm0}
\\
\delta_{box}C_{1q}(m=0) =& 
- \frac{\alpha}{4\pi} Q_{e}Q_{q} \cdot 6 g_{VA}^{eq} 
\left(\ln \left(\frac{M_Z^2}{M^2}\right) + \frac{3}{2}\right)
\, , 
\label{C1thresholdm0}
\\
\delta_{box}C_{2q}(m=0) =& 
- \frac{\alpha}{4\pi} Q_{e}Q_{q} \cdot 6 g_{AV}^{eq} 
\left(\ln \left(\frac{M_Z^2}{M^2}\right) + \frac{5}{6}\right)
\, , 
\label{C2thresholdm0}
\\
\delta_{box}C_{3q}(m=0) =& 
\frac{\alpha}{4\pi} Q_{e}Q_{q} \cdot 8 g_{VV}^{eq} 
\left(\ln \left(\frac{M_Z^2}{M^2}\right) + 1\right)
\, ; 
\label{C3thresholdm0}
\end{align}
and for zero quark mass: 
\begin{align}
\delta_{box}C_{0q}(M=0) =& 
\frac{\alpha}{4\pi} Q_{e}Q_{q} \cdot 6 g_{AA}^{eq} 
 \left(\ln \left(\frac{M_Z^2}{m^2}\right) + \frac{3}{2}\right)
\, , 
\label{C0thresholdM0}
\\
\delta_{box}C_{1q}(M=0) =& 
- \frac{\alpha}{4\pi} Q_{e}Q_{q} \cdot 6 g_{VA}^{eq} 
 \left(\ln \left(\frac{M_Z^2}{m^2}\right) + \frac{5}{6}\right)
\, , 
\label{C1thresholdM0}
\\
\delta_{box}C_{2q}(M=0) =& 
- \frac{\alpha}{4\pi} Q_{e}Q_{q} \cdot 6 g_{AV}^{eq} 
\left(\ln \left(\frac{M_Z^2}{m^2}\right) + \frac{3}{2}\right)
\, , 
\label{C2thresholdM0}
\\
\delta_{box}C_{3q}(M=0) =& 
\frac{\alpha}{4\pi} Q_{e}Q_{q} \cdot 8 g_{VV}^{eq} 
 \left(\ln \left(\frac{M_Z^2}{m^2}\right) + 1\right)
\, . 
\label{C3thresholdM0}
\end{align}
It is interesting to see how the pattern of vector- and axial-vector 
couplings simplifies: while in the generic case with both masses 
non-zero, there is always a contribution from both the `direct' 
and the `inverted' combination of $V$ and $A$-couplings, i.e.\ 
for example the correction to $C_{1q} = g_{AV}^{eq}$ has terms 
proportional to both $g_{AV}^{eq}$ and $g_{VA}^{eq}$, but after 
taking one of the masses to zero, only the 'inverted' combination 
survives\footnote{
   Terms proportional to $g_{AV}^{eq}$ and $g_{VA}^{eq}$ 
   contributing to the box-graph correction have been called 
   `vector' and `axial-vector' box in the literature, $\square_{\gamma Z}^{V}$ 
   and $\square_{\gamma Z}^{A}$, resp.}. 

\begin{figure}[t!]
\centering
\begin{subfigure}[b]{0.48\textwidth}
    \includegraphics[width=\textwidth]{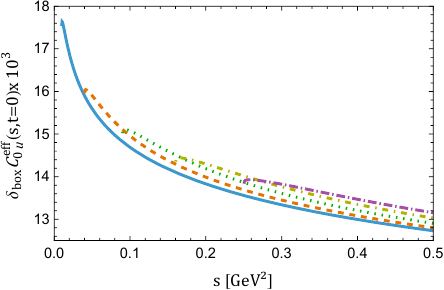}
    \caption{$\delta_{box}C_{0u}^{\mathrm{eff}}(s,t=0)$}
    \vspace*{3mm}
    \label{C0MassScale}
\end{subfigure}
%\hspace{0.5 cm}
\begin{subfigure}[b]{0.49\textwidth}
    \includegraphics[width=\textwidth]{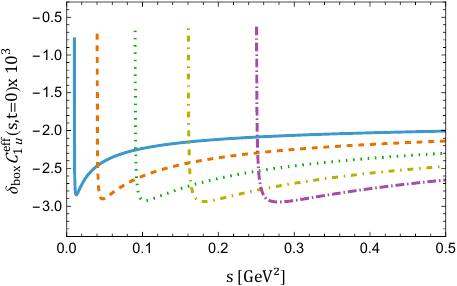}
    \caption{$\delta_{box}C_{1u}^{\mathrm{eff}}(s,t=0)$}
    \vspace*{3mm}
    \label{C1MassScale}
\end{subfigure}
\begin{subfigure}[b]{0.49\textwidth}
    \includegraphics[width=\textwidth]{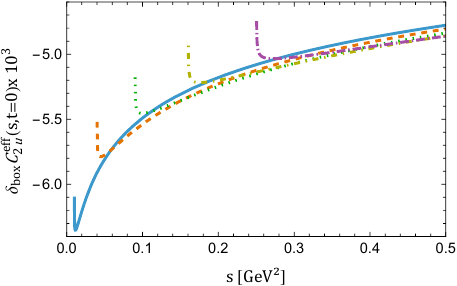}
    \caption{$\delta_{box}C_{2u}^{\mathrm{eff}}(s,t=0)$}
    \label{C2MassScale}
\end{subfigure}
\begin{subfigure}[b]{0.47\textwidth}
    \includegraphics[width=\textwidth]{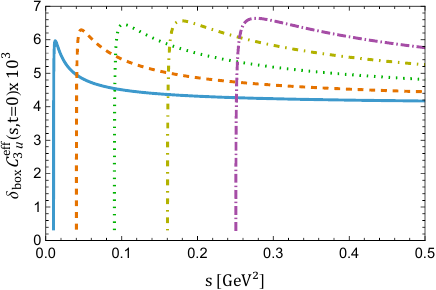}
    \caption{$\delta_{box}C_{3u}^{\mathrm{eff}}(s,t=0)$}
    \label{C3MassScale}
\end{subfigure}
\caption{ 
The box-graph corrections $\delta_{box}C_{i u}^{\mathrm{eff}}$ as a function 
of the Mandelstam variable $s$ at fixed $t=0$ for different effective quark 
masses $M$: blue for  $0.1~\mathrm{GeV}$, orange for $0.2~\mathrm{GeV}$, 
green for $0.3~\mathrm{GeV}$, yellow for $0.4~\mathrm{GeV}$, and purple for 
$0.5~\mathrm{GeV}$.}
\label{MscaleC0C3}
\end{figure}

Compared with the results found in the previous literature, 
Refs.~\cite{Marciano:1982mm,Erler:2013xha}, we observe 
that there is indeed a large logarithm, but the constants are 
different for the different effective couplings. Most importantly, 
we see that the argument of the logarithm depends on the 
assumption which of the two masses should be taken zero. 
If one insisted that a non-zero quark mass should be used to 
regularize the singularity, there will be an ambiguity from the 
choice of the numerical value of the quark mass. We show a 
few choices in Fig.~\ref{MscaleC0C3}. However, it is in fact 
possible to use zero quark mass, thereby removing all 
sensitivity to low-scale hadronic physics. 

%%%%%%%%%%%%%%%%%%%%%%%%
%%%%%%%%%%%%%%%%%%%%%%%%

At high energies, for deep-inelastic scattering, it is well-known 
that the logarithmic, singular quark mass dependence can be 
factorized and absorbed into parton distribution functions, PDFs. 
The PDFs can be defined universally because at high energies 
there are no mass logarithms from the box graphs (see 
Appendix~\ref{app:largeE}). 
The quark mass in a calculation of box graphs can therefore 
be taken zero. Otherwise there would be logarithms of the quark 
mass with coefficients that depend on the product of quark and 
electron couplings, e.g.\ the correction $\delta_{box}C_{1q}$ is 
proportional to $g_V^e g_A^q$. Such correction terms cannot 
be factorized in a universal way and, therefore, cannot be 
absorbed into parton distribution functions. 
We argue that a similar approach should eventually be possible 
also at low energies in such a way that the dependence on 
the internal structure of the target is separated from the 
perturbative result.

We end this section by combining our results into the box-graph 
correction which contributes to the observable polarization 
asymmetry, as introduced at the end of section~\ref{sec:notation}. 
The box graphs are both $s$-  and $t$-dependent and there are 
additional pre-factors from contractions of the spin vector with 
4-momenta, see Eq.~(\ref{epolarized}). These kinematic factors 
have to be taken into account if we were to calculate the asymmetry 
for scattering off a single heavy fermion, and we have 
\begin{align}
\square_{\gamma Z}^{eq}(s,t) 
= 
& 
- 2 \delta_{box}C_{1q}^{\text{eff}}(s,t) \times
\nonumber \\ \quad \quad \quad 
& \quad
\frac{ 2 \lambda(s, m^2, M^2) (s - m^2 - M^2) + 2 s (s - (m^2 + M^2) ) t + (s + m^2 - M^2) t^2}{
Q_e Q_q \sqrt{\lambda(s, m^2, M^2)} \left( 2 (s - (m^2 + M^2) )^2 + 2 s t + t^2 \right)} 
\nonumber \\[1ex] 
&
+ 2 \delta_{box}C_{2q}^{\text{eff}}(s,t) \times
\nonumber \\ \quad \quad \quad 
& \quad
\frac{ 2 ((m^2 - M^2)^2 - 2 (m^2 + M^2) s + s^2) t + (s + m^2 - M^2) t^2 }{
Q_e Q_q \sqrt{\lambda(s, m^2, M^2)}
\left( 2 (s - (m^2 + M^2))^2 + 2 s t + t^2 \right)} 
\label{eq:deltaBoxeq}
\end{align}
with the triangle (K\"all\'en) function 
\begin{equation} 
\lambda(x,y,z) = x^2 + y^2 + z^2 - 2 x y -2 x z - 2 y z \, . 
\end{equation} 
The denominator in Eq.~(\ref{eq:deltaBoxeq}) stems from the 
normalization to the unpolarized cross section, i.e.\ the 
leading-order one-photon exchange. The correction 
to the $C_2$-coefficient contributes for non-forward scattering, 
i.e.\ if $t \neq 0$. 
This expression would be 
needed for example for the scattering off muons, with 
corresponding replacements for the couplings to muons 
instead of quarks. For zero momentum transfer, $t = 0$, 
this simplifies to 
\begin{align}
\square_{\gamma Z}^{eq}(s,t=0) 
= 
&
- 2 \delta_{box}C_{1q}^{\text{eff}}(s,0) 
\frac{ \sqrt{ \lambda(s, m^2, M^2)}  }{Q_e Q_q \left(s - m^2 - M^2\right)} \, . 
\label{eq:deltaBoxAPV}
\end{align}
We note that in the reference frame where the heavy fermion with 
mass $M$ is at rest, the momentum of the incoming electron, 
$p_1$ can be written as  
\begin{equation} 
p_1 = \frac{1}{2M} \sqrt{(s - (M+m)^2) (s - (M-m)^2)} = 
\frac{\sqrt{ \lambda(s, m^2, M^2)} }{2M}\, , 
\end{equation} 
and therefore 
\begin{align}
\square_{\gamma Z}^{eq}(s,t=0) 
= 
& 
- 2 \frac{ p_1 }{E_1} \,  
\, \delta_{box}C_{1q}^{\text{eff}}(s,0) 
\end{align}
with $E_1 = \sqrt{p_1^2 + m^2}$. For the energy at threshold, which 
depends on the masses, the result vanishes due to the factor $p_1$. 
At larger energies there is only a weak residual $s$-dependence and the 
$\gamma Z$-box-graph correction amounts to about $6 - 7 \times 10^{-3}$. 
Note that at $s$ above threshold both vector and axial-vector components 
contribute, i.e.\ terms proportional to $g_{VA}^{eq}$ and $g_{AV}^{eq}$, resp. 
Graphs for the $s$-dependence of the two components look very similar, but 
the axial-vector component is larger than the vector component by a factor 
of roughly 4. 

The results above are indeed also required to describe deep inelastic 
electron scattering off a nucleon. In this case, cross sections are 
obtained from summing electron-quark scattering cross sections 
weighted with parton distribution functions. At energies and momentum 
transfers well below the weak scale, $s, Q^2 \ll M_Z^2$, the threshold 
limit Eq.~(\ref{eq:threshold}) may provide an approximation which is 
precise enough. We expect that such an approach is valid for example 
for previous measurements at JLab \cite{Wang:2014guo}, or even for 
future measurements at the EIC \cite{AbdulKhalek:2021gbh}. 

%%%%%%%%%%%%%%%%%%%%%%%%
\begin{figure}[b!]
\centering 
       \begin{picture}(300,240)(0,0)
       \put(0,0){\includegraphics[width=0.7\textwidth]{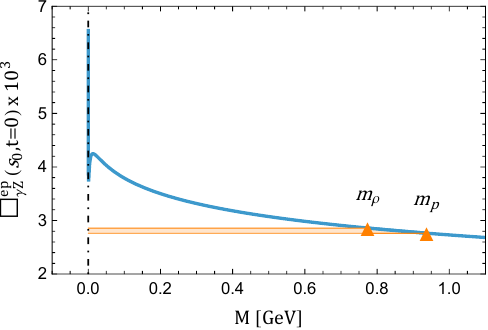}} 
       \put(165,110){\includegraphics[width=0.33\textwidth]{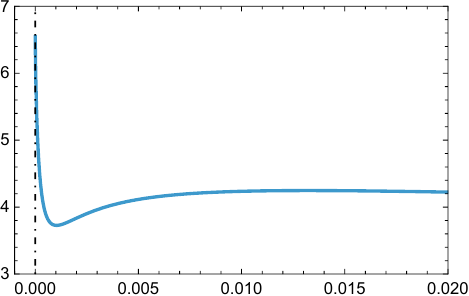}} 
       \end{picture}
\caption{ 
The constant term of the box-graph correction $\Box_{\gamma Z}^{ep}$ 
at threshold contributing to the parity-violating asymmetry $A_{PV}$ for 
scattering of an electron off a proton from Eq.~(\ref{eq:deltaBox-ep}) 
using the general case Eq.~(\ref{C1thresholdgeneric}), as a function of 
the quark mass. The horizontal orange band covers the  range for 
quark masses between the $\rho$-mass and the proton-mass. 
The region at small $M$ where the very small value 
of the electron mass determines the shape of the function 
is shown at an increased scale in the zoom-in plot. 
}
\label{fig:APV_ep_M}
\end{figure}
%%%%%%%%%%%%%%%%%%%%%%%%

For elastic electron proton scattering, only the $s$-independent 
threshold values of $\delta_{box}C_{1q}$ in Eqs.~(\ref{C1thresholdm0}, 
\ref{C1thresholdM0}) are needed since only these constants can be 
absorbed into the low-energy effective Lagrangian. We are now in a 
position to rewrite Eq.~(\ref{eq:gammaZboxES}): instead of using an 
ambiguous hadronic cutoff ($\Lambda$ in Eq.~(\ref{eq:gammaZboxES}), 
or an effective quark mass $M$ in our perturbative calculation), one 
can separate the complete $\gamma Z$-box-graph correction in a term 
where the quark mass is set to zero plus an energy-dependent 
remainder $F_{\gamma Z}^{ep}(s,t)$. 
%which contains all the information 
%about the non-perturbative hadronic structure in terms of form factors: 
%which contains information about the non-perturbative hadronic structure.
This function contains the  hadronic dynamics absent from the free-quark calculation and therefore accounts for the transition from the effective electron--quark description to the physical proton observable. 

\begin{equation}
\square_{\gamma Z}^{ep}(s,t) 
= 
- 2 \left(2\delta_{box}C_{1u}(s_0,0) + \delta_{box}C_{1d}(s_0,0) \right) 
+ F_{\gamma Z}^{ep}(s,t) 
\, , 
\label{eq:deltaBox-ep}
\end{equation}
Any dependence of the low-energy couplings on the nucleon structure 
is avoided in this way by setting the quark mass to zero. In this limit, for the low-energy coupling constant 
term we find 
\begin{equation}
- 2 \left(2\delta_{box}C_{1u}(s_0,0) + \delta_{box}C_{1d}(s_0,0) \right) 
= 6.55 \times 10^{-3} \quad \text{for} ~ M=0 \, . 
\end{equation}
Had we chosen to keep a non-zero quark mass, for example 
$M = 0.3$~GeV, the correction would be different, 
\begin{equation}
- 2 \left(2\delta_{box}C_{1u}(s_0,0) + \delta_{box}C_{1d}(s_0,0) \right) 
= 3.39 \times 10^{-3}  \quad \text{for} ~ M=0.3~ \text{GeV} \, . 
\end{equation} 
Details of the quark-mass dependence of this quantity is shown in 
Fig.~\ref{fig:APV_ep_M}. The complicated structure of the graph at 
small mass values is due to the interplay of quark- and electron-mass 
values.  
We emphasize that different choices for the quark mass correspond 
to different ways to separate perturbative from non-perturbative 
effects in the calculation. A shift of the quark mass has to correspond 
to a modification of the non-perturbative part $F_{\gamma Z}^{ep}(s,t)$. 

A correct implementation of corrections from the $\gamma Z$-box 
graphs is extremely important for the interpretation of polarized 
electron-proton scattering since the weak charge of the proton is 
very small and therefore the relative correction large. The situation 
is less severe for the extraction of the weak mixing angle from 
atomic parity violation in heavy nuclei. If $F^{ep}_{\gamma Z}(s_0,0)$ term is neglected, the weak charge of a nucleus 
containing $Z$ protons and $N$ neutrons is, at leading order, 
\begin{equation}
Q_W^{Z,N} = - 2 (Z g_{AV}^{ep} + N g_{AV}^{en})
\end{equation} 
with 
\begin{align}
g_{AV}^{ep} &= 2 g_{AV}^{eu} + g_{AV}^{ed} \equiv 2 C_{1u} + C_{1d} \, , \\
g_{AV}^{en} &= g_{AV}^{eu} + 2 g_{AV}^{ed} \equiv C_{1u} + 2 C_{1d} \, 
\end{align}
For the important case of ${}^{133}$Cs where precise experimental 
results exist, Ref.~\cite{Sahoo:2021thl} found $Q_W^{55,78} = 
-73.71(26)_{ex}(23)_{th}$ including corrections. Using either 
Eq.~(\ref{C1thresholdm0}) with the $\rho$-mass as a regulator, 
or our new result, Eq.~(\ref{C1thresholdM0}) with zero quark mass, 
we observe a shift of the nucleus weak charge by $-0.43$. 
Extracting the weak mixing angle based on one or the other prescription 
for the $\gamma Z$-box, this would lead to a shift of $\Delta s_W^2 = 
0.0020$, to be compared with the uncertainty of $\pm 0.0016$ for 
$s_W^2$ found in Ref.~\cite{Sahoo:2021thl}. Also in this case, the 
treatment of $\gamma Z$-box-graph corrections can therefore lead 
to significant effects. 

We do not discuss the $t$-dependence of our results since for non-zero 
$t$ there are infrared divergent terms which have to be cancelled against 
contributions from real photon radiation. For soft photon radiation an 
analytic calculation is possible, neglecting the energy of the photon. 
This, however, requires introducing a soft-photon cutoff which makes 
numerical results ambiguous. Only after taking into account also hard 
photon radiation, the bremsstrahlung correction will be well-defined.

%%%%%%%%%%%%%%%%%%%%%%%%%%%%%%%%%%%%%%%%%%%%%%%%%%%%%
\section{Conclusions}
\label{sec:conclusion}

Our study of the Standard Model one-loop result for the 
$\gamma Z$-box graphs has revealed that a careful treatment 
of the fermion masses is needed to obtain a well-defined limit at zero 
momentum transfer and at threshold energy. We have calculated  
complete analytic expressions without any approximation for 
the one-loop box graphs and obtain simple formulae for the 
threshold limit. With both masses non-zero, there is always a 
contribution from both the direct and the inverted combination 
of V and A-couplings. In particular, we found that low-energy couplings 
are well-defined when the quark mass is set to zero at the end 
of the calculation. The box-graph contributions to the low-energy 
couplings shown in Eqs.~(\ref{C0thresholdM0} - \ref{C3thresholdM0})
are the main result of this calculation and should replace what 
was used previously in the literature. 
For a full energy-dependent determination of the $\gamma Z$-box correction, these results have to be consistently combined with the corresponding non-perturbative contribution, the precise implementation of which remains to be worked out.

Our results for the low-energy couplings with zero quark mass 
are free of uncertainties from the hadron structure. In an 
application to polarized electron nucleus scattering, 
hadron structure dependence is then contained in form factors 
and additional, probably sub-leading, hadronic matrix elements 
of operators which go beyond the leading current matrix 
elements of the one-boson exchange approximation. Predictions 
for these terms  
are certainly needed for a reliable 
interpretation of corresponding measurements at Qweak 
and P2@MESA. This will require different techniques, however, 
beyond a naive perturbative calculation, for example using 
dispersion relations or lattice QCD and goes beyond the scope 
of the present work. In addition, also Coulomb corrections 
should be studied. From our perturbative one-loop calculation 
we can not exclude the possibility that such corrections are 
significant at energies where experimental data are available 
or will come up in the near future.  

Apart from the impact of our study on the determination of 
the weak charge of the proton, $\gamma Z$-box-graph 
corrections are also needed in other applications. We 
have mentioned atomic parity violation above. Another 
case could be the analysis of deep inelastic scattering 
data. Both the extraction of Standard Model parameters, 
most notably the weak mixing angle, as well as of limits on 
parameters of beyond the Standard Model physics may be 
affected. Corresponding investigations are left to future work.

%\clearpage

%%%%%%%%%%%%%%%%%%%%%%%%%%%%%%%%%%%%%%%%%%%%%%%%%%%%%
\begin{appendix}
%%%%%%%%%%%%%%%%%%%%%%%%%%%%%%%%%%%%%%%%%%%%%%%%%%%%%

%%%%%%%%%%%%%%%%%%%%%%%%%%%%%%%%%%%%%%%%%%%%%%%%%%%%%
\section{$s$-dependent result in the forward limit}
\label{sec:sdependence}

In the following section, we present the leading-order corrections, 
$\delta_{box} \hat{C}_i(s,t=0)$, to the effective couplings as functions 
of the initial energy $s$, evaluated in the strict forward-scattering limit 
($t = 0$). Although the method is applicable to arbitrary kinematic 
configurations, we focus here exclusively on the forward-scattering 
case due to the complexity of the full expressions. We use the 
abbreviation $u_0 = 2(m^2 + M^2) - s$. 
\begin{align} 
& 
\delta_{\text{box}} \hat{C}_0 (s,t=0) 
=\;   
\frac{\alpha}{4 \pi} Q_{e}Q_{q}
\frac{M_Z^2}{m^2 + M^2 - s} \;  \textup{Re} 
\nonumber \\ & \quad 
\Bigg( 
g_{VV}^{eq} 
\bigg[
    -8 \text{B}_0(0,0,M_Z^2) 
  +4 \text{B}_0(m^2,0,m^2) 
  +4 \text{B}_0(m^2,m^2,M_Z^2) 
  +4\text{B}_0(M^2,0,M^2)  
\nonumber \\ & \quad 
  \left. 
  +4 \text{B}_0(M^2,M^2,M_Z^2) 
  -4 \text{B}_0(u_0, m^2, M^2)   
  -4\text{B}_0(s, m^2, M^2) \right. 
\nonumber \\ & \quad
  \left.
  +  4\left(4 m^2 + M_Z^2\right) 
    \text{C}_0(m^2,0,m^2,m^2,M_Z^2,0) +4\left(4 M^2 + M_Z^2\right) 
    \text{C}_0(M^2,0,M^2,M^2,M_Z^2,0) \right. 
\nonumber \\ & \quad
  \left.
   -2(4 m^2 + 4 M^2 + M_Z^2 - 2 s) 
    \text{C}_0(M^2,u_0,m^2,0,M^2,m^2) \right. 
\nonumber \\ & \quad
  \left.
   -2(4 m^2 + 4 M^2 + M_Z^2 - 2 s) 
    \text{C}_0(M^2,u_0,m^2,M_Z^2,M^2,m^2)\right. 
\nonumber \\ & \quad
  \left.
  -2(M_Z^2 + 2s)
    \text{C}_0(m^2,M^2,s,m^2,0,M^2) 
   -2(M_Z^2 + 2s) 
    \text{C}_0(M^2,s,m^2,M_Z^2,M^2,m^2) \right. 
\nonumber \\ & \quad
  -2[2 M_{Z}^2(2(m^2+M^2)-s)+4(m^2+M^2-s)^2+M_Z^4] 
  \text{D}_0(m^2,M^2,M^2,m^2, u_0, 0,m^2,M_Z^2,M^2,0)
\nonumber \\ & \quad
  -2[4(m^2+M^2-s)^2+2 s M_Z^2+M_Z^4] \text{D}_0(m^2,M^2,M^2,m^2,s,0,m^2,M_Z^2,M^2,0)
\bigg] 
\nonumber \\ & \quad
+ g_{AA}^{eq} 
\bigg[
 -4 \text{B}_0(u_0, m^2, M^2)
  +4 \text{B}_0(s, m^2, M^2) 
\nonumber \\  & \quad
  +\frac{2[-2m^4+M_Z^2(m^2+M^2-s)+4m^2(M^2+s)-2(M^2-s)^2]}{m^2+M^2-s} 
  \text{C}_0(m^2,M^2,s,m^2,0,M^2) 
\nonumber \\ & \quad
  +\frac{2[M_Z^2(m^2+M^2-s)-2(m^4+m^2(6M^2-2s)+(M^2-s)^2)]}{m^2+M^2-s} 
  \text{C}_0(M^2,s,m^2,M_Z^2,M^2,m^2)  
\nonumber \\ & \quad
- \frac{2[M_Z^2(m^2+M^2-s)+2(m^4-2m^2(M^2+s)+(M^2-s)^2)]}
    {m^2 + M^2 - s} \text{C}_0(M^2,u_0,m^2,0,M^2,m^2) 
\nonumber\\ & \quad
  - \frac{2[M_Z^2(m^2+M^2-s)+2(m^4+m^2(6M^2-2s)+(M^2-s)^2]}
    {m^2 + M^2 - s} \text{C}_0(M^2,u_0,m^2,M_Z^2,M^2,m^2)
\nonumber \\ & \quad
  + \frac{2\left(2((m+M)^2-s)((m-M)^2-s) M_Z^2 - (m^2 + M^2 - s) M_Z^4 \right)}
    {m^2 + M^2 - s} 
\nonumber \\ & \quad \quad 
    \times \text{D}_0(m^2,M^2,M^2,m^2,s,0,m^2,M_Z^2,M^2,0) 
\nonumber \\ & \quad \left.
  -  \frac{2\left((m^2 + M^2 - s) M_Z^4 + 2(m^4-2m^2(M^2+s)+(M^2-s)^2) M_Z^2 
  \right)}{m^2 + M^2 - s} \right. 
\nonumber \\ & \quad \quad 
  \times \text{D}_0(m^2,M^2,M^2,m^2,u_0,0,m^2,M_Z^2,M^2,0) 
\bigg]
\Bigg) \, , 
\label{eq:deltaboxC0st}
\end{align}

\begin{align} 
& 
\delta_{\text{box}} \hat{C}_1 (s,t=0) = 
\frac{\alpha}{4\pi} \, Q_{e}Q_{q} 
\frac{-M_Z^2}{2 M^2((m-M)^2-s)((m+M)^2-s)}
 \textup{Re} 
\notag \\ & \quad  
\Bigg(
g_{VA}^{eq}
\bigg[
   + 16 m^2 M^2 \text{B}_0\left(m^2,0,m^2\right)
   + 16 m^2 M^2 \text{B}_0\left(m^2,m^2,M_Z^2\right)
\notag \\ & \quad \quad \quad 
   - 8 M^2 \left(3m^2+M^2-s\right) \text{B}_0\left(u_0, m^2,M^2\right) 
   - 8 M^2 \left(m^2-M^2+s\right) \text{B}_0\left(s,m^2,M^2\right) 
\notag \\& \quad \quad \quad  
   +16 m^2 M^2 M_Z^2 \text{C}_0\left(m^2,0,m^2,m^2,M_Z^2,0\right) 
\notag \\ & \quad \quad \quad  
   + \left(M^4-\left(m^2-s\right)^2\right) M_Z^2 \text{C}_0\left(m^2,M^2,s,m^2,0,M^2\right) 
\notag \\ & \quad \quad \quad 
   - \left(m^2+M^2-s\right) \left(m^2+3 M^2-s\right) M_Z^2 
     \text{C}_0\left(M^2, u_0, m^2,0,M^2,m^2\right) 
\notag \\ & \quad \quad \quad 
   + \bigg(\left(m^4-8 M^2 m^2-2 s m^2+7 M^4+s^2-8 M^2 s\right) M_Z^2  
\notag \\ & \quad \quad \quad \quad 
   - 16 M^2 \left(m^4-2 \left(M^2+s\right) m^2+\left(M^2-s\right)^2\right) \bigg) 
     \text{C}_0\left(M^2,s,m^2,M_Z^2,M^2,m^2\right)
\notag \\ & \quad \quad \quad 
   + \bigg(\left(m^4-20 M^2 m^2-2 s m^2-5 M^4+s^2+4 M^2 s\right) M_Z^2 
\notag \\ & \quad \quad \quad \quad 
   -16 M^2 \left(m^4-2 \left(M^2+s\right) m^2+\left(M^2-s\right)^2\right) \bigg) 
    \text{C}_0\left(M^2, u_0, m^2,M_Z^2,M^2,m^2\right) &\quad \quad \quad \quad - M_Z^4 \left(m^2+M^2-s\right) \left(m^2+3 M^2-s\right)
   \text{D}_0\left(m^2,M^2,M^2,m^2, u_0, 0,m^2,M_Z^2,M^2,0\right) 
\notag \\ & \quad \quad \quad 
   + M_Z^4 \left(m^2+M^2-s\right) \left(-m^2+M^2+s\right) 
     \text{D}_0\left(m^2,M^2,M^2,m^2,s,0,m^2,M_Z^2,M^2,0\right) \bigg]
\notag \\  & \quad  
   + g_{AV}^{eq} 
\bigg[
      8 M^2 \left(m^2+M^2-s\right) \text{B}_0\left(M^2,0,M^2\right) 
   + 8 M^2 \left(m^2+M^2-s\right) \text{B}_0\left(M^2,M^2,M_Z^2\right)  
\notag \\ & \quad \quad \quad 
   - 8 M^2 \left(3 m^2+M^2-s\right) \text{B}_0\left( u_0, m^2,M^2\right) 
   + 8 M^2 \left(m^2-M^2+s\right) \text{B}_0\left(s,m^2,M^2\right) 
\notag \\ & \quad \quad \quad 
   + 32 m^2 M^2 \left(m^2+M^2-s\right) \text{C}_0\left(m^2,0,m^2,m^2,M_Z^2,0\right) 
\notag \\ & \quad \quad \quad 
   + 8 M^2 \left(m^2+M^2-s\right) \left(4 M^2+M_Z^2\right) 
   \text{C}_0\left(M^2,M^2,0,0,M^2,M_Z^2\right) 
\notag \\ & \quad \quad \quad 
   - \left(\left(m^2-s\right)^2-M^4\right)  \left(2 m^2-2M^2-M_Z^2-2 s\right) 
   \text{C}_0\left(m^2,M^2,s,m^2,0,M^2\right) 
\notag \\ & \quad \quad \quad 
   - \left(m^2+M^2-s\right) \left(m^2+3 M^2-s\right) \left(2 m^2+6 M^2+M_Z^2-2 s\right) 
   \text{C}_0\left(M^2, u_0, m^2,0,M^2,m^2\right) 
\notag \\ & \quad \quad \quad 
   - \left(-\left(\left(m^4-20 M^2 m^2-2 s m^2-5 M^4+s^2+4 M^2
   s\right) M_Z^2\right) \right. 
\notag \\ & \quad \quad \quad \quad 
   \left. 
   -2 \left(m^2+M^2-s\right) \left(m^4-2 \left(5 M^2+s\right) m^2-7 M^4+s^2+2 M^2 s\right)
   \right) 
   \text{C}_0\left(M^2, u_0, m^2,M_Z^2,M^2,m^2\right) 
\notag \\ & \quad \quad \quad 
   - \left(\left(m^4-2 \left(4 M^2+s\right) m^2+7 M^4+s^2-8 M^2 s\right) M_Z^2 
   \right. 
\notag \\ & \quad \quad \quad \quad 
   \left. 
      - 2 \left(m^2+M^2-s\right) \left(m^4-2 \left(M^2+s\right) m^2+M^4+s^2-6 M^2 s\right)\right) 
   \text{C}_0\left(M^2,s,m^2,M_Z^2,M^2,m^2\right)
\notag \\ & \quad \quad \quad 
   - \left(m^2+M^2-s\right) 
      \left(\left(m^2+3 M^2-s\right) M_Z^4+2 \left(m^2+3M^2-s\right)^2 M_Z^2
      \right. 
\notag \\ & \quad \quad \quad \quad 
   \left. 
   +16 M^2 \left((m-M)^2-s\right) \left((m+M)^2-s\right)\right) 
     \text{D}_0\left(m^2,M^2,M^2,m^2, u_0, 0,m^2,M_Z^2,M^2,0\right) 
\notag \\ & \quad \quad \quad 
   - \left(m^2+M^2-s\right)
     \left(\left(-m^2+M^2+s\right) M_Z^4+2\left(-m^2+M^2+s\right)^2 M_Z^2 
     \right. 
\notag \\ & \quad \quad \quad \quad 
   \left. 
   + 16 M^2 \left((m-M)^2-s\right) \left((m+M)^2-s\right)
   \right) 
     \text{D}_0\left(m^2,M^2,M^2,m^2,s,0,m^2,M_Z^2,M^2,0\right) 
\bigg] 
\Bigg) \, , 
\label{eq:deltaboxC1st}
\end{align}

\begin{align} 
&
\delta_{\text{box}} \hat{C}_2 (s,t=0) = \; 
\frac{\alpha}{4\pi} Q_{e}Q_{q} 
\frac{- M_Z^2}{2 m^2((m-M)^2-s)((m+M)^2-s)} 
\textup{Re} 
\notag \\ & \quad  
\Bigg( 
g_{AV}^{eq} 
\bigg[
  + 16 m^2 M^2 \text{B}_0(M^2, 0, M^2) 
  + 16 m^2 M^2 \text{B}_0(M^2, M^2, M_Z^2) 
\notag \\ & \quad  \quad \quad 
  - 8 m^2 (m^2 + 3 M^2 - s) \text{B}_0(u_0, m^2, M^2) 
  + 8 m^2 (m^2 - M^2 - s) \text{B}_0(s, m^2, M^2) 
\notag \\ & \quad  \quad \quad 
  + (m^2 + M^2 - s)(m^2 - M^2 + s) M_Z^2 \text{C}_0(m^2, M^2, s, m^2, 0, M^2)  
\notag \\ & \quad  \quad \quad 
  + 16 m^2 M^2 M_Z^2 \text{C}_0(M^2, M^2, 0, 0, M^2, M_Z^2)  
\notag \\ & \quad  \quad \quad 
  - (m^2 + M^2 - s)(3 m^2 + M^2 - s) M_Z^2 \text{C}_0(M^2, u_0, m^2, 0, M^2, m^2)  
\notag \\ & \quad  \quad \quad 
  - \Big[ 
     16 (m^4 - 2 (M^2 + s) m^2 + (M^2 - s)^2) m^2 
     + (-7 m^4 + 8 (M^2 + s) m^2 - (M^2 - s)^2)  
    \Big] 
\notag \\ & \quad  \quad \quad \quad 
   \times \text{C}_0(M^2, s, m^2, M_Z^2, M^2, m^2) 
\notag \\ & \quad  \quad \quad 
  - \Big[
    (5 m^4 + 4(5 M^2 - s) m^2 - (M^2 - s)^2) M_Z^2 
   + 16 m^2 (m^4 - 2 (M^2 + s) m^2 + (M^2 - s)^2)  
    \Big] 
\notag \\ & \quad  \quad \quad \quad 
     \times \text{C}_0(M^2, u_0, m^2, M_Z^2, M^2, m^2) 
\notag \\ &  \quad \quad \quad 
  - (m^2 + M^2 - s)(3 m^2 + M^2 - s) 
  \text{D}_0(m^2, M^2, M^2, m^2, u_0, 0, m^2, M_Z^2, M^2, 0) M_Z^4 
\notag \\ & \quad \quad \quad 
  + (m^2 + M^2 - s)(m^2 - M^2 + s) 
  \text{D}_0(m^2, M^2, M^2, m^2, s, 0, m^2, M_Z^2, M^2, 0) M_Z^4 
\bigg] 
\notag \\ & \quad 
+ g_{VA}^{eq} 
\bigg[
  8 m^2 (m^2 + M^2 - s) \left[ \text{B}_0(m^2, 0, m^2) + \text{B}_0(m^2, m^2, M_Z^2) \right] 
\notag \\ & \quad  \quad \quad 
  - 8 m^2 (m^2 + 3 M^2 - s) \text{B}_0(u_0, m^2, M^2) 
  - 8 m^2 (m^2 - M^2 - s) \text{B}_0(s, m^2, M^2) 
\notag \\ & \quad  \quad \quad 
  + 32 m^2 M^2 (m^2 + M^2 - s) \text{C}_0(M^2, M^2, 0, 0, M^2, M_Z^2) 
\notag \\ & \quad  \quad \quad 
  + (m^2 + M^2 - s)(m^2 - M^2 + s) (-M_Z^2 - 2(m^2 - M^2 + s)) 
    \text{C}_0(m^2, M^2, s, m^2, 0, M^2)    
\notag \\ & \quad  \quad \quad 
  + 8 m^2 (m^2 + M^2 - s) (4 m^2 + M_Z^2) \text{C}_0(m^2, 0, m^2, m^2, M_Z^2, 0) 
\notag \\ & \quad  \quad \quad 
  - (m^2 + M^2 - s)(3 m^2 + M^2 - s)  (6 m^2 + 2 M^2 + M_Z^2 - 2 s) 
      \text{C}_0(M^2, u_0, m^2, 0, M^2, m^2) 
\notag \\ & \quad  \quad \quad 
+ \Big[
    2 (m^2 + M^2 - s)(m^4 - 2 (M^2 + 3 s) m^2 + (M^2 - s)^2) 
%\notag \\ & \quad  \quad \quad  
  - (7 m^4 - 8(M^2 + s) m^2 + (M^2 - s)^2) M_Z^2 
\Big] 
\notag \\ & \quad  \quad \quad \quad
  \times \text{C}_0(M^2, s, m^2, M_Z^2, M^2, m^2) 
\notag \\ & \quad  \quad \quad 
  + \left[-2 (7 m^4 + 2(5 M^2 - s)m^2 - (M^2 - s)^2)(m^2 + M^2 - s) \right. 
\notag \\ &  \quad \quad \quad \quad
     \left. - (5 m^4 + 4(5 M^2 - s)m^2 - (M^2 - s)^2) M_Z^2 \right]
  \text{C}_0(M^2, u_0, m^2, M_Z^2, M^2, m^2) 
\notag \\ & \quad  \quad \quad 
  + (m^2 + M^2 - s) 
\notag \\ & \quad  \quad \quad \quad 
  \times \left[
    - (3 m^2 + M^2 - s) M_Z^4 
    - 2 (3 m^2 + M^2 - s)^2 M_Z^2 
    - 16 m^2 ((m - M)^2 - s)((m + M)^2 - s)
  \right] 
\notag \\ & \quad  \quad \quad \quad
  \times \text{D}_0(m^2, M^2, M^2, m^2, u_0, 0, m^2, M_Z^2, M^2, 0) 
\notag \\ & \quad  \quad \quad 
  + (m^2 + M^2 - s) 
\notag \\ & \quad  \quad \quad \quad 
  \times \left[
    - (m^2 - M^2 + s) M_Z^4 
    - 2 (m^2 - M^2 + s)^2 M_Z^2 
    - 16 m^2 ((m - M)^2 - s)((m + M)^2 - s)
\right] 
\notag \\ & \quad  \quad \quad \quad
  \times \text{D}_0(m^2, M^2, M^2, m^2, s, 0, m^2, M_Z^2, M^2, 0) 
\bigg]
\Bigg) \, ,  
\label{eq:deltaboxC2st}
\end{align}

\begin{align} 
&
\delta_{\text{box}} \hat{C}_3 (s,t=0) = 
\frac{\alpha}{4 \pi} Q_{e}Q_{q} \frac{M_Z^2}{m^2 + M^2 - s} 
\;  \textup{Re} 
\notag \\ & \quad
\Bigg( 
g_{AA}^{eq} 
  \bigg[
    -2 \left(2(m^2 + M^2) - M_Z^2 \right) 
    \text{C}_0(m^2, M^2, s, m^2, 0, M^2) \notag \\
  & \quad - 2 \left(2(m^2 + M^2) - M_Z^2 \right) 
    \text{C}_0(M^2, s, m^2, M_Z^2, M^2, m^2) \notag \\
  & \quad - 2(4(m^2+M^2-s)^2-2(m^2+M^2)M_Z^2+M_Z^4) \notag \\
  & \quad \times \text{D}_0(m^2, M^2, M^2, m^2, s, 0, m^2, M_Z^2, M^2, 0) \notag\\
  & \quad + 4(m^2 + M^2) 
    \text{C}_0(M^2, u_{0}, m^2, 0, M^2, m^2) \notag \\
  & \quad - 4(m^2 + M^2) 
    \text{C}_0(M^2, u_{0}, m^2, M_Z^2, M^2, m^2) \notag \\
  & \quad 
    + 4  \left[(m^2 + M^2) M_Z^2 - 2 (m^2 + M^2 - s)^2 \right]  \notag \\
  & \qquad \times \text{D}_0(m^2, M^2, M^2, m^2, u_0, 0, m^2, M_Z^2, M^2, 0)
\bigg] 
\notag \\ &
g_{VV}^{eq}
\bigg[
    2 \left(M_Z^2 - 2(m^2 + M^2 - s) \right)
    \text{C}_0(m^2, M^2, s, m^2, 0, M^2) \notag \\
  & \quad - 2 \left(2(m^2 + M^2 - s) + M_Z^2 \right)
    \text{C}_0(M^2, s, m^2, M_Z^2, M^2, m^2) \notag \\
  & \quad + 2 M_{Z}^2[M_{Z}^2-2(m^2+M^2-s)] \notag \\
  & \quad \times \text{D}_0(m^2, M^2, M^2, m^2, s, 0, m^2, M_Z^2, M^2, 0) \notag \\
  & \quad - 4(m^2+M^2-s)
    \text{C}_0(M^2, u_{0}, m^2, 0, M^2, m^2) \notag \\
  & \quad - 4 (m^2+M^2-s) 
    \text{C}_0(M^2, u_{0}, m^2, M_Z^2, M^2, m^2) \notag \\
  & \quad - 4 (m^2+M^2-s)M_{Z}^2 
    \text{D}_0(m^2, M^2, M^2, m^2, u_0, 0, m^2, M_Z^2, M^2, 0)
\bigg]
\Bigg) \, .
\label{eq:deltaboxC3st}
\end{align}

%%%%%%%%%%%%%%%%%%%%%%%%%%%%%%%%%%%%%%%%%%%%%%%%%%%%%
\section{$\gamma Z$-box at large energy}
\label{app:largeE}

For completeness we present in this appendix also a result 
for the $\gamma Z$ box graphs which is valid at large energy 
and momentum transfer. Here we use the notation of 
Ref.~\cite{Bohm:1986rj}, see App.~B.4, Eq.~(B.13) 
there\footnote{Note that in Ref.~\cite{Bohm:1986rj} the factor 
  $1/(2 s_W c_W)$ was absorbed into the coupling constants 
  $v_f$, $a_f$.}. 
Since there are no mass singularities remaining after combining 
the direct and crossed box graphs, it is possible to take the limit 
where the fermion masses are zero. We correct two typos of 
Ref.~\cite{Bohm:1986rj}: (1) the overall coefficient $1/s$ has to 
be replaced by $1/(s-M_Z^2)$; (2) the expression for the integral 
$I^{\gamma Z}$ has to be completely anti-symmetrized with 
respect to exchanging $t$ and $u$. The complete expression 
is given here for the case of electron-quark scattering (i.e.\ in 
the $t$-channel). The incoming 4-momenta of the electron (quark) 
are denoted $p^\mu$ ($q^\mu$) and those of the outgoing 
particles correspondingly $p^{\prime\mu}$ ($q^{\prime\mu}$). 
The matrix element can be written in terms of two integrals 
$I^{\gamma Z}$ and $I_5^{\gamma Z}$ which can be expressed 
as functions of the Mandelstam invariants 
\begin{eqnarray} 
s = (p+q)^2, \quad \quad 
t = (p - p^\prime)^2, \quad \quad 
u = (p - q^\prime)^2 
\label{eq:stu}
\end{eqnarray}
with $s + t + u = 0$, i.e.\ terms suppressed with powers of the 
fermion masses, $m_f^2/s$, $m_f^2/t$, $m_f^2/u$ are omitted. 
The sum of the direct and crossed box graphs for electron-quark 
scattering is 
\begin{eqnarray}
B^{\gamma Z}_{ef} 
&=& 
\left(\frac{\alpha}{2\pi}\right)^2 
\frac{Q_e Q_f}{4 c_W^2 s_W^2} 
\Bigl\{
\bar{u}(p^\prime) 
\gamma_\mu \left(g_V^e - g_A^e \gamma_5\right) 
u(p) \cdot 
\bar{u}(q^\prime) 
\gamma^\mu \left(g_V^f - g_A^f \gamma_5\right) 
u(q) 
\nonumber
\\[1ex] \nonumber
&& \quad\quad\quad\quad\quad\quad
\times \left(I^{\gamma Z}(t,u) - I^{\gamma Z}(t,s) \right)
\\[1ex] 
\nonumber
&& \quad\quad\quad\quad\quad 
+ \,\, \bar{u}(p^\prime) 
\gamma_\mu \left(g_A^e - g_V^e \gamma_5\right) 
u(p) \cdot 
\bar{u}(q^\prime) 
\gamma^\mu \left(g_A^f - g_V^f \gamma_5\right) 
u(q) 
\\[1ex] 
&& \quad\quad\quad\quad\quad\quad
\times 
\left(I_5^{\gamma Z}(t,u) + I_5^{\gamma Z}(t,s\right)
\Bigr\}. 
\label{eq:m2box}
\end{eqnarray}
This can be sorted like the low-energy effective Lagrangian 
used above, see Eq.~(\ref{eq:lowEL})
\begin{eqnarray}
B^{\gamma Z}_{ef} 
&=& 
\tilde{C}_{0f}
\bar{u}(p^\prime) \gamma_\mu u(p) \cdot 
\bar{u}(q^\prime) \gamma^\mu u(q) 
+ 
\tilde{C}_{1f}
\bar{u}(p^\prime) \gamma_\mu \gamma_5 u(p) \cdot 
\bar{u}(q^\prime) \gamma^\mu u(q) 
\nonumber
\\[1ex] \nonumber
&& 
+ \,\, 
\tilde{C}_{2f}
\bar{u}(p^\prime) \gamma_\mu u(p) \cdot 
\bar{u}(q^\prime) \gamma^\mu \gamma_5 u(q) 
+ 
\tilde{C}_{3f}
\bar{u}(p^\prime) \gamma_\mu \gamma_5 u(p) \cdot 
\bar{u}(q^\prime) \gamma^\mu \gamma_5 u(q) 
\label{eq:lowBEL}
\end{eqnarray}
with 
\begin{eqnarray}
\tilde{C}_{0f}
&=& 
\left(\frac{\alpha}{2\pi}\right)^2 
\frac{Q_e Q_f}{4 c_W^2 s_W^2} 
\left[ 
g_V^e g_V^f \left(I^{\gamma Z}(t,u) - I^{\gamma Z}(t,s) \right) 
+ g_A^e g_A^f \left(I_5^{\gamma Z}(t,u) + I_5^{\gamma Z}(t,s) \right) 
\right] \, , 
\\[1ex] \nonumber
\tilde{C}_{1f}
&=& 
- \left(\frac{\alpha}{2\pi}\right)^2 
\frac{Q_e Q_f}{4 c_W^2 s_W^2} 
\left[ 
g_A^e g_V^f \left(I^{\gamma Z}(t,u) - I^{\gamma Z}(t,s) \right) 
+ g_V^e g_A^f \left(I_5^{\gamma Z}(t,u) + I_5^{\gamma Z}(t,s) \right) 
\right] \, , 
\\[1ex] \nonumber
\tilde{C}_{2f}
&=& 
- \left(\frac{\alpha}{2\pi}\right)^2 
\frac{Q_e Q_f}{4 c_W^2 s_W^2} 
\left[ 
g_V^e g_A^f \left(I^{\gamma Z}(t,u) - I^{\gamma Z}(t,s) \right) 
+ g_A^e g_V^f \left(I_5^{\gamma Z}(t,u) + I_5^{\gamma Z}(t,s) \right) 
\right] \, , 
\\[1ex] \nonumber
\tilde{C}_{3f}
&=& 
\left(\frac{\alpha}{2\pi}\right)^2 
\frac{Q_e Q_f}{4 c_W^2 s_W^2} 
\left[ 
g_A^e g_A^f \left(I^{\gamma Z}(t,u) - I^{\gamma Z}(t,s) \right) 
+ g_V^e g_V^f \left(I_5^{\gamma Z}(t,u) + I_5^{\gamma Z}(t,s) \right) 
\right] \, . 
\end{eqnarray}
The crossed box is contained in the above expressions. 
It was obtained from the direct box by exchanging 
$s \leftrightarrow u$ and changing the sign of $I^{\gamma Z}$. 
For electron-antiquark scattering (or positron-quark scattering) 
one has to exchange $s$ with $u$, i.e.\ in the sum of the direct 
and the crossed box, terms which contain $I^{\gamma Z}$ 
($I_5^{\gamma Z}$) are odd (even) with respect to exchanging 
one of the particles by its anti-particle. The result is ultraviolet 
finite and reads 
\begin{eqnarray}
\label{eq:I5}
I_5^{\gamma Z}(t,u) 
&=& 
\frac{1}{2(t+u)} 
\Biggl\{
\ln\left(\frac{u}{t-M_Z^2}\right)
- \frac{M_Z^2}{t}\ln\left(\frac{M_Z^2}{M_Z^2-t}\right)
\\[1ex] && 
\hspace{20mm}
+ \frac{t + 2u + M_Z^2}{t+u} 
\Biggl[
{\rm Li}_2\left(\frac{t}{M_Z^2}\right)
- {\rm Li}_2\left(\frac{t+u}{M_Z^2}\right)
\nonumber 
\\[1ex] && 
\hspace{48mm}
+ \ln\left(\frac{t+u}{M_Z^2}\right)
  \ln\left(\frac{M_Z^2-t}{M_Z^2-t-u}\right) 
\Biggr]
\Biggr\} ,
\nonumber
\end{eqnarray}
\begin{eqnarray}
\label{eq:Ia}
I^{\gamma Z}(t,u) - I^{\gamma Z}(t,s) 
&=& I_5^{\gamma Z}(t,u) - I_5^{\gamma Z}(t,s) 
\\[1ex] &&
+ \frac{1}{t-M_Z^2} 
\Biggl\{
{\rm Li}_2\left(\frac{M_Z^2+s}{s}\right)
- {\rm Li}_2\left(\frac{M_Z^2+u}{u}\right) 
\nonumber 
\\[1ex] && 
\hspace{18mm}
+ \frac{1}{2} \ln\left(\frac{s^2}{u^2}\right)
  \ln\left(\frac{M_Z^2}{M_Z^2-t}\right)
+ \ln\left(\frac{-u}{s}\right)\ln\left(\frac{M_Z^2}{\lambda^2}\right)
\Biggr\}.
\nonumber
\end{eqnarray} 
The last term in Eq.~(\ref{eq:Ia}) contains the infrared 
divergence which was regularized with the help of a finite 
photon mass $\lambda$. The result agrees with 
Ref.~\cite{Brown:1983jv} (after correcting a missing factor 
of 2 for a term containing $\ln(u/t) \ln(1 - s/M_Z^2)$) and with 
Ref.~\cite{Fujimoto:1990tb}. 

For the case of deep inelastic scattering the kinematic 
variables can be written in terms of the dimensionless 
scaling variables $x$ and $y$: 
\begin{eqnarray} 
s = x S, \quad \quad 
t = - xy S, \quad \quad 
u = - x(1-y) S, 
\label{eq:DISxy}
\end{eqnarray}
where $S=(p+P)^2$ is the squared center-of-mass energy of the 
electron-nucleon initial state\footnote{Note that also the 
  nucleon mass should be neglected for DIS, in concordance 
  with the assumption that electron-nucleon scattering is  
  described by electron-quark scattering in the infinite 
  momentum frame.} 
and the physical range is simply given by 
\begin{equation}
0 < x < 1, \quad \quad 
0 < y < 1. 
\end{equation} 
Introducing also the abbreviation 
\begin{equation}
\mu_Z^2 = \frac{M_Z^2}{S} ,
\label{eq:mratio} 
\end{equation} 
the box-graph integrals can be written in terms of dimensionless 
variables only: 

\begin{eqnarray}
\label{eq:I5x,y}
S I_5^{\gamma Z}(t,s) 
&=& 
- \frac{1}{2x(1-y)} 
\Biggl\{
\ln\left(\frac{x}{xy + \mu_Z^2}\right)
+ \frac{\mu_Z^2}{xy}\ln\left(\frac{\mu_Z^2}{xy + \mu_Z^2}\right)
\\[1ex] && 
\hspace{24mm}
+ \frac{x + x(1-y) + \mu_Z^2}{x(1-y)} 
\Biggl[
{\rm Li}_2\left(\frac{-xy}{\mu_Z^2}\right)
- {\rm Li}_2\left(\frac{-x}{\mu_Z^2}\right)
\nonumber 
\\[1ex] && 
\hspace{62mm}
+ \ln\left(\frac{x}{\mu_Z^2}\right)
  \ln\left(\frac{xy + \mu_Z^2}{x + \mu_Z^2}\right) 
\Biggr]
\Biggr\} ,
\nonumber 
%%%
\\[2ex]
S I_5^{\gamma Z}(t,u) 
&=& 
\frac{1}{2x} 
\Biggl\{
\ln\left(\frac{x(1-y)}{xy + \mu_Z^2}\right)
+ \frac{\mu_Z^2}{xy}\ln\left(\frac{\mu_Z^2}{xy + \mu_Z^2}\right)
\\[1ex] && 
\hspace{10mm}
+ \frac{x + x(1-y) - \mu_Z^2}{x} 
\Biggl[
{\rm Li}_2\left(\frac{-xy}{\mu_Z^2}\right)
- {\rm Li}_2\left(\frac{x(1-y)}{\mu_Z^2} + i\epsilon \right)
\nonumber 
\\[1ex] && 
\hspace{50mm}
+ \ln\left(\frac{x(1-y)}{\mu_Z^2}\right)
  \ln\left(\frac{xy + \mu_Z^2}{-x(1-y) + \mu_Z^2}\right) 
\Biggr]
\Biggr\} ,
\nonumber 
\end{eqnarray}

\begin{eqnarray}
\label{eq:Iaxy}
S\left(I^{\gamma Z}(t,s) - I^{\gamma Z}(t,u)\right) 
&=& 
S\left(I_5^{\gamma Z}(t,s) - I_5^{\gamma Z}(t,u)\right) 
\\[1ex] &&
- 
\frac{1}{xy + \mu_Z^2}
\Biggl\{
{\rm Li}_2\left(\frac{x+\mu_Z^2}{x}\right)
- {\rm Li}_2\left(\frac{x(1-y)+\mu_Z^2}{x(1-y)}\right) 
\nonumber 
\\[1ex] && 
\hspace{20mm}
- \ln\left(1-y\right)
  \ln\left(\frac{\mu_Z^2}{xy+\mu_Z^2}\right)
+ \ln\left(1-y\right)\ln\left(\frac{M_Z^2}{\lambda^2}\right)
\Biggr\}.
\nonumber
\end{eqnarray}
This form is particularly helpful to determine the limit 
for vanishing momentum transfer, i.e.\ $y \rightarrow 0$ 
and the low-energy / high-mass limit is obtained by 
$\mu_Z^2 \rightarrow \infty$. From the last equation, 
Eq.~(\ref{eq:Iaxy}) one can read off that the anti-symmetric 
parts of the two box integrals become equal to each other and 
the infrared divergence becomes suppressed by $y$ (i.e.\ $t$). 
It turns out that also the soft-photon bremsstrahlung contribution 
has this property, i.e.\ vanishes with $t \rightarrow 0$ 
and the contribution of the $\gamma Z$-box graphs to the 
low-energy effective couplings is completely determined 
by the $y \rightarrow 0$, $\mu_Z^2 \rightarrow \infty$ limit 
of the above expressions. 
%I find (from Mathematica, {\tt BoxGZ-xy-v6.nb}): 
We find 
\begin{eqnarray}
I(0) &=& M_Z^2 \lim_{\mu_Z^2\to\infty} \lim_{y\to 0} 
       \left( I^{\gamma Z}(t,u) - I^{\gamma Z}(t,s)  
       \right) 
     = O(1/\mu_Z^2) + O(y) , 
\\ 
I_5(0) &=& M_Z^2 \lim_{\mu_Z^2\to\infty} \lim_{y\to 0} 
       \left( I_5^{\gamma Z}(t,u) + I_5^{\gamma Z}(t,s)  
       \right) 
     = 
     - \frac{3}{2} 
     \left(\ln\left(\frac{M_Z^2}{s}\right) + \frac{7}{6} \right) . 
\end{eqnarray} 
Normalizing to $G_F / \sqrt{2}$, the correction of the low-energy 
effective couplings are then
\begin{eqnarray}
\delta_{box}C_{1q}^{\text{eff}}(s,t=0)
&=& \frac{\alpha}{2\pi} Q_e Q_q \, g_{VA}^{eq} \, 
    3 \left(\ln\left(\frac{M_Z^2}{s}\right) + \frac{7}{6} \right) ,
\\ 
\delta_{box}C_{2q}^{\text{eff}}(s,t=0)
&=& \frac{\alpha}{2\pi} Q_e Q_q \, g_{AV}^{eq} \, 
    3 \left(\ln\left(\frac{M_Z^2}{s}\right) + \frac{7}{6} \right) . 
\end{eqnarray} 
This looks similar to the results of \cite{Marciano:1982mm}, 
but disagrees in two respects: (1) the logarithm contains the 
center-of-mass energy $s$, not an ill-defined hadronic mass 
scale, and (2) the constant is $\frac{7}{6}$, not $\frac{3}{2}$ 
or $\frac{5}{6}$, relative to the logarithm. Note also that the 
correction factor is the same for the two couplings, in contrast 
to what we found above. However, since fermion masses have 
been neglected, the low-energy limit, i.e.\ $s \to 0$ can not be 
considered the correct final result for the corrections we 
are looking for; the logarithm of $s$ is divergent for $s \to 0$.

%%%%%%%%%%%%%%%%%%%%%%%%%%%%%%%%%%%%%%%%%%%%%%%%%%%%%
\end{appendix}
%%%%%%%%%%%%%%%%%%%%%%%%%%%%%%%%%%%%%%%%%%%%%%%%%%%%%

\section*{Acknowledgment} 
We thank Yannick Ulrich and Sophie Kollatzsch for discussions and for providing 
us with a Mathematica notebook containing results for the $\gamma Z$ box 
graphs of their $s$-channel calculation for $e^+ e^- \to \mu^+ \mu^-$ in 
Ref.~\cite{Kollatzsch:2022bqa}. We acknowledge helpful discussions with 
Jens Erler and Mikhail Gorchtein, as well as with Chien-Yeah Seng who 
confirmed the presence of the Sommerfeld enhancement term by an independent 
calculation. 
This work has been supported by the Cluster of Excellence ``Precision Physics, Fundamental Interactions, and Structure of Matter'' (PRISMA++ EXC 2118/2), funded by the German Research Foundation (DFG) within the German Excellence Strategy (Project ID 390831469). H.~S. has received funding from the European Research Council (ERC) under the European Union's Horizon research and innovation programme (Grant Agreement No.~101142600). The work of B.~D. and P.~M. has been supported by the Ministerio de Ciencia e Innovación under grant PID2023-146142NB-I00, by the Secretaria d'Universitats i Recerca del Departament d'Empresa i Coneixement de la Generalitat de Catalunya, and by the Spanish Ministry of Science and Innovation (MICINN) through the State Research Agency under the Severo Ochoa Centres of Excellence Programme 2025--2029 (CEX2024-001442-S). IFAE is partially funded by the CERCA programme of the Generalitat de Catalunya. B.~D. also acknowledges support from the predoctoral programme AGAUR-FI (2025 FI-1 00461) Joan Oró of the Department of Research and Universities of the Generalitat de Catalunya, co-financed by the European Social Fund Plus.

%%%%%%%%%%%%%%%%%%%%%%%%%%%%%%%%%%%%%%%%%%%%%%%%%%%%%
%%%%%%%%%%%%%%%%%%%%%%%%%%%%%%%%%%%%%%%%%%%%%%%%%%%%%

%%%%%%%%%%%%%%%%%%%%%%%%%%%%%%%%%%%%%%%%%%%%%%%%%%%%%
%%%%%%%%%%%%%%%%%%%%%%%%%%%%%%%%%%%%%%%%%%%%%%%%%%%%%

\end{document}